\documentclass[prd,twocolumn,floatfix,amsmath,nofootinbib,amssymb,floatfix]{revtex4}
\usepackage{graphicx,color,dcolumn,booktabs,bm}
\usepackage{longtable,lscape}
\usepackage{pdfpages}
\usepackage{txfonts}
\usepackage{overpic}
\usepackage{amssymb}
\usepackage{makecell}
\usepackage{indentfirst}
\usepackage{feynmf}   
\usepackage{slashed}  
\usepackage{cases}
\usepackage{color}
\usepackage{multirow}
\usepackage{threeparttable}
\usepackage{epstopdf}
\usepackage{enumerate}
\usepackage{subfigure}
\usepackage{diagbox}
\usepackage{graphicx,color,dcolumn,booktabs,bm}
\usepackage{mathrsfs}
\usepackage{cancel}
\usepackage{float}
\usepackage[misc]{ifsym}
\usepackage[colorlinks,
            citecolor=blue,
            anchorcolor=red,
            menucolor=red,
            linkcolor=red,
            filecolor=red,
            runcolor=red,
            urlcolor=blue,
            frenchlinks=red]{hyperref}
\usepackage{array}
\usepackage{booktabs}

\begin{document}
\title{Systematic analysis of the form factors of $B_{c}$ to $P$-wave charmonia and corresponding weak decays}
\author{Jie Lu$^{1,2}$}
\email{l17693567997@163.com}
\author{Dian-Yong Chen$^{1,3}$}
\email{chendy@seu.edu.cn}
\author{Guo-Liang Yu$^{2}$}
\email{yuguoliang2011@163.com}
\author{Zhi-Gang Wang$^{2}$}
\email{zgwang@aliyun.com}
\author{Bin Wu$^{2}$}

\affiliation{$^1$ School of Physics, Southeast University, Nanjing 210094, People's Republic of
China\\$^2$ Department of Mathematics and Physics, North China
Electric Power University, Baoding 071003, People's Republic of
China\\$^3$ Lanzhou Center for Theoretical Physics, Lanzhou University, Lanzhou 730000, People's Republic of
China}
\date{\today }

\begin{abstract}
In this article, the vector, axial vector and tensor form factors of $B_{c}\to \chi_{cJ}$ ($J=0,1,2$) and $B_{c}\to h_{c}$ are analyzed within the framework of three-point QCD sum rules. With the calculated vector and axial vector form factors, we directly study the decay widths and branching ratios of semileptonic decays $B_{c}^{-}\to \chi_{cJ}l \bar{\nu}_l, h_{c}l \bar{\nu}_l$ $(l=e, \mu$ and $\tau)$ and analyze the nonleptonic decays $B_{c}^{-}\to \chi_{cJ}\pi^{-}, \chi_{cJ}K^{-}, \chi_{cJ}\rho^{-}, \chi_{cJ}K^{*-}$, $B_{c}^{-}\to h_{c}\pi^{-}, h_{c}K^{-}, h_{c}\rho^{-}, h_{c}K^{*-}$ by using the naive factorization approach (NFA). These results can provide more information to understand the properties of $B_{c}$ meson and $P$-wave charmonia and to study the heavy quark dynamics.
\end{abstract}

\pacs{13.25.Ft; 14.40.Lb}

\maketitle

\section{Introduction}\label{sec1}

Since the $B_{c}$ meson was firstly observed by CDF collaboration in 1998~\cite{CDF:1998axz}, most of theoretical and experimental physicists payed much attention to it for the special properties. Firstly, as the only quarkonium with mixed heavy flavor quarks, $B_{c}$ meson has rich decay channels because each heavy quark in it can decay individually and the other acts as a spectator. Secondly, the mass of $B_{c}$ meson lies below the threshold of $B\bar{D}$, thus the $B_{c}$ meson can only decay via weak interaction. This provides a suitable place to test the standard model (SM) and find the new physics beyond the SM. Thirdly, both $b$ and $c$ quarks can annihilate and provide new kinds of weak decays with sizable partial decay widths. These pure leptonic or radiative leptonic decay can be used to extract the decay constant of $B_{c}$ and the Cabibbo-Kobayashi-Maskawa (CKM) matrix element $V_{cb}$~\cite{Chang:1992pt,Chang:1997re}. In addition, the $B_{c}$ meson includes two different heavy flavor quarks, the spectroscopy may be different with the light mesons or the mesons with only one heavy quark. It offers us a different laboratory to study the strong interactions. Moreover, it was estimated that the inclusive production cross sections of $B_{c}$ meson and its resonance states at the LHC is at level of 1 $\mathrm{\mu b}$ for $\sqrt{s}=\sqrt{14}$ TeV. This means that there are about $O(10^{9})$ $B_{c}$ meson can be anticipated in 1 fb$^{-1}$~\cite{Gao:2010zzc}. So, abundant events in experiment can provide more information to carry out this research.

The weak decays of $B_{c}$ to charmonia including semileptonic and nonleptonic processes are interesting and valuable. Taking $B_{c}^{-}$ meson as an example,  its weak decay processes can be described as $b$ quark decaying while the other $\bar{c}$ quark playing as a spectator at the quark level. In the case of semileptonic decays, the decay mode of $b$ quark is $b\to cl\bar{\nu}_{l}$ ($l=e, \mu$ and $\tau$). The final $c$ quark will combine with the spectator $\bar{c}$ quark to produce final charmonium. In the nonleptonic decay processes, the decay modes of $b$ quark include $b\to c\bar{u}d$, $c\bar{u}s$, $cd\bar{c}$ and $cs\bar{c}$. Except for the final $c$ quark that combines with spectator $\bar{c}$ quark to produce final charmonium, the rest quarks combine with each other to produce another final meson. Theoretical investigations about these decay processes are difficult because the Quantum Chromodynamics (QCD) is non-perturbative in low energy regions. Therefore, some non-perturbative methods have been employed to study the decay processes of $B_{c}$ to charmonia in recent years, such as the lattice QCD (LQCD)~\cite{Colquhoun:2016osw,Harrison:2020gvo}, the QCD sum rules (QCDSR)~\cite{Colangelo:1992cx,Kiselev:2000pp,Azizi:2009ny,Azizi:2013zta,Wu:2024gcq}, the light-cone QCD sum rules (LCSR)~\cite{Huang:2007kb,Wang:2007fs,Leljak:2019eyw,Bordone:2022drp}, the perturbative QCD (pQCD)~\cite{Kiselev:2001zb,Rui:2017pre,Liu:2018kuo,Liu:2020upy}, the relativistic quark
model (RQM)~\cite{Nobes:2000pm,Ivanov:2006ni,Ebert:2010zu}, the non-relativistic quark model (NRQM)~\cite{Hernandez:2006gt}, the light-front quark model (LFQM)~\cite{Wang:2008xt,Wang:2009mi,Zhang:2023ypl,Li:2023wgq,Hazra:2023zdk,S:2024adt}, non-relativistic QCD (NRQCD)~\cite{Qiao:2012hp,Zhu:2017lwi,Tang:2022nqm} and other methods~\cite{Chang:2001pm,Tran:2018kuv}. With the updated high luminosity of the LHC, more experimental measurements of $B_{c}$ to charmoina will become feasible in the coming years, this will provide more information to test various theoretical models.

As one of the most powerful non-perturbative approach, the QCD sum rules are widely used in studying the properties of hadrons such as the masses, decay constants and light-cone distribution amplitude of hadrons~\cite{Wei:2006wa,Wang:2013ff,Chen:2015kpa,Wang:2015mxa,Chen:2016phw,Zhang:2021wnv,Wang:2022xja,Huang:2022xny,Su:2024lzy,Wang:2024fwc}, the form factors of hadron transition~\cite{Wang:2007ys,Wang:2008pq,Shi:2019hbf,Peng:2020ivn,Zhao:2020mod,Neishabouri:2024gbc,Tousi:2024usi,Lu:2024tgy}, the coupling constants of strong interaction~\cite{Bracco:2011pg,Wang:2013iia,Azizi:2014bua,Azizi:2015tya,Yu:2015xwa,Rodrigues:2017qsm,Yu:2019sqp,Lu:2023gmd,Lu:2023pcg} and others~\cite{Wang:2017dce,Dehghan:2023ytx,Ozdem:2024lpk}. In our previous work, the scalar, vector, axial vector and tensor form factors of $B_{c}$ to $S$-wave charmonia $\eta_{c}$ and $J/\psi$ were analyzed in the framework of three-point QCD sum rules, and the corresponding semileptonic and nonleptonic decays processes were studied~\cite{Wu:2024gcq}. As a continuation of our previous work, we systematically analyze the vector, axial vector and tensor form factors of $B_{c}$ to $P$-wave charmonia $\chi_{cJ} (J=0, 1, 2)$ and $h_{c}$ by using three-point QCD sum rules in the present work. With these form factors, we also study the semileptonic decay processes $B_{c}^{-}\to \chi_{cJ}l \bar{\nu}_l, h_{c}l \bar{\nu}_l$ $(l=e, \mu$ and $\tau)$ and the nonleptonic decay processes $B_{c}^{-}\to \chi_{cJ}\pi^{-}, \chi_{cJ}K^{-}, \chi_{cJ}\rho^{-}, \chi_{cJ}K^{*-}$, $B_{c}^{-}\to h_{c}\pi^{-}, h_{c}K^{-}, h_{c}\rho^{-}, h_{c}K^{*-}$. The nonleptonic decay $B_{c}\to \chi_{cJ}D_{(s)}^{(*)}$ and $h_{c}D_{(s)}^{(*)}$ are also related to the form factors in this study. However, other form factors involved in these decay processes, are beyond the scope of the present work.   

This work is organized as follows. After introduction, the form factors of $B_{c}$ to $P$-wave charmonia are analyzed by three-point QCD sum rules in Sec. \ref{sec2}. Based on these form factors, we analyze the corresponding semileptonic and nonleptonic decay widths and branching ratios in Sec. \ref{sec3}. Sec. \ref{sec4} is employed to present the numerical results and discussions and Sec. \ref{sec5} is devoted to a short summary. Some important figures and formulas are shown in Appendixes \ref{Sec:AppA}$\sim$\ref{Sec:AppC}.

\section{Three-point QCD sum rules for transition form factors}\label{sec2}

To study the form factors of $B_{c}$ to $P$-wave charmonia, we firstly write down the following three-point correlation function,
\begin{eqnarray}\label{eq:1}
\notag
\Pi (p,p') &&= i^2\int d^4xd^4ye^{ip'x}e^{i(p - p')y} \\
&& \times \left\langle 0 \right| T\{ J_X(x)\tilde J(y)J_{B_c}^+(0)\} \left| 0 \right\rangle 
\end{eqnarray}
where $T$ denotes the time ordered product, $J_{B_{c}}$ and $J_{X}$ ($X=\chi_{cJ}$ and $h_{c}$, $J=0,1,2$) are the interpolating currents of $B_{c}$ meson and $P$-wave charmonia, respectively.  $\tilde{J}$ is the transition currents. These currents are taken as following forms,
\begin{eqnarray}\label{eq:2}
\notag
J_{B_c}(0) &&= \bar c(0)i\gamma _5b(0)\\
\notag
\tilde J(y) &&= \bar c(y)\Gamma b(y)\\
\notag
J^{\chi _{c0}}(x) &&= \bar c(x)c(x)\\
\notag
J_\mu ^{\chi _{c1}}(x) &&= \bar c(x)\gamma _\mu \gamma _5 c(x)\\
\notag
J_{\mu \nu }^{h_c}(x) &&= \bar c(x)\sigma _{\mu \nu }c(x)\\
J_{\mu \nu }^{\chi _{c2}}(x) &&= \frac{i}{2}\bar c(x)({\gamma _\mu }{\mathord{\buildrel{\lower3pt\hbox{$\scriptscriptstyle\leftrightarrow$}} 
\over D} }_\nu + {\gamma _\nu }{\mathord{\buildrel{\lower3pt\hbox{$\scriptscriptstyle\leftrightarrow$}} 
\over D} }_\mu-\frac{1}{2}g_{\mu\nu}{\mathord{\buildrel{\lower3pt\hbox{$\scriptscriptstyle\leftrightarrow$}} 
\over{\slashed D}}})c(x)
\end{eqnarray}
where $\Gamma=\gamma_{\mu}, \gamma_{\mu}\gamma_{5}$ and $\sigma_{\mu\nu}$ or $\sigma_{\mu\nu}\gamma_{5}$ for vector, axial vector and tensor form factors, respectively. $\mathord{\buildrel{\lower3pt\hbox{$\scriptscriptstyle\leftrightarrow$}} 
\over D}_\mu = \frac{1}{2}(\mathord{\buildrel{\lower3pt\hbox{$\scriptscriptstyle\rightarrow$}} 
\over D}_\mu - \mathord{\buildrel{\lower3pt\hbox{$\scriptscriptstyle\leftarrow$}} 
\over D}_\mu )$, $\mathord{\buildrel{\lower3pt\hbox{$\scriptscriptstyle\rightarrow$}} 
\over D}_\mu = \mathord{\buildrel{\lower3pt\hbox{$\scriptscriptstyle\rightarrow$}} 
\over \partial} _\mu - ig_st^aG^a_\mu$ and $\mathord{\buildrel{\lower3pt\hbox{$\scriptscriptstyle\leftarrow$}} 
\over D}_\mu = \mathord{\buildrel{\lower3pt\hbox{$\scriptscriptstyle\leftarrow$}} 
\over \partial} _\mu + ig_st^aG^a_\mu$ denote the covariant derivative.  

In the framework of QCD sum rules, the correlation function will be calculated at both hadron and quark levels which are called phenomenological side and QCD side, respectively. Combining these two sides by using quark hadron duality, the sum rules for the form factors will be derived.

\subsection{The phenomenological side}\label{sec2.1}
In phenomenological side, complete sets of hadronic states with the same quantum numbers as interpolating currents are inserted in correlation function in Eq. (\ref{eq:1}). Performing the integration in coordinate space and using the double dispersion relation, the correlation function can be written as the following form~\cite{Colangelo:2000dp},
\begin{eqnarray}\label{eq:3}
\notag
\Pi ^{\mathrm{phy}}(p,p') &&= \frac{\left\langle 0 \right|J_X(0)\left| X(p') \right\rangle \left\langle B_c(p) \right|J_{B_c}^ + (0)\left| 0 \right\rangle }{(p'^2 - m_X^2)(p^2 - m_{B_c}^2)}\\
&& \times \left\langle X(p') \right|\tilde J(0)\left| B_c(p) \right\rangle  + ...
\end{eqnarray}
where the ellipsis represent the contributions of higher resonances and continuum states. The meson vacuum matrix elements in above correlation function are defined as, 
\begin{eqnarray}\label{eq:4}
\notag
\left\langle 0 \right|J^{\chi _{c0}}(0)\left| \chi _{c0}(p') \right\rangle  &&= f_{\chi _{c0}}m_{\chi _{c0}}\\
\notag
\left\langle 0 \right|J_\mu ^{\chi _{c1}}(0)\left| \chi _{c1}(p') \right\rangle  &&= f_{\chi _{c1}}m_{\chi _{c1}}\epsilon^{\chi_{c1}}_\mu \\
\notag
\left\langle 0 \right|J_\mu ^{\chi _{c1}}(0)\left| {\eta _c}(p') \right\rangle  &&= if_{\eta _c}p'_\mu \\
\notag
\left\langle 0 \right|J_{\mu \nu }^{h_c}(0)\left| h_c(p') \right\rangle  &&= if_{h_c}\varepsilon _{\mu \nu \alpha \beta }\epsilon^{h_c}_\alpha p'_\beta \\
\notag
\left\langle 0 \right|J_{\mu \nu }^{h_c}(0)\left| J/\psi (p') \right\rangle  &&= if_{J/\psi}(p'_\mu \epsilon^{J/\psi}_\nu - p'_\nu \epsilon^{J/\psi}_\mu )\\
\notag
\left\langle 0 \right|J_{\mu \nu} ^{\chi _{c2}}(0)\left|\chi _{c2}(p') \right\rangle  &&= f_{\chi _{c2}}m_{\chi _{c2}}^3\epsilon^{\chi_{c2}}_{\mu \nu} \\
\left\langle B_c(p) \right|J_{B_c}^ + (0)\left| 0 \right\rangle  &&= \frac{f_{B_c}m_{B_c}^2}{m_b + m_c}
\end{eqnarray}
where $f_{\chi_{c0}}$, $f_{\chi_{c1}}$, $f_{\eta_{c}}$, $f_{h_{c}}$, $f_{J/\psi}$, $f_{\chi_{c2}}$ and $f_{B_{c}}$ are the decay constants of corresponding mesons,
$\epsilon^{\chi_{c1}}_{\mu}$, $\epsilon^{h_c}_{\alpha}$ and $\epsilon^{J/\psi}_{\mu}$ are the polarization vectors of $\chi_{c1}$, $h_{c}$ and $J/\psi$, respectively. $\epsilon^{\chi_{c2}}_{\mu\nu}$ is the polarization tensor of $\chi_{c2}$, and $\varepsilon_{\mu\nu\alpha\beta}$ is the 4-dimension Levi-Civita tensor. It is noted that the current of $\chi_{c1}$ ($J^{P}=1^{+}$) can couple with the pseudoscalar charmonium $\eta_{c}$ ($J^{P}=0^{-}$). The tensor current $\bar{c}\sigma_{\mu\nu}c$ which is selected to interpolate $h_{c}$ ($J^{P}=1^{+}$) can also couple with $J/\psi$ ($J^{P}=1^{-}$). In order to eliminate the contamination of the redundant coupling, the projection operator should be induced in corresponding correlation function which will be discussed in next sections. The transition matrix elements in Eq. (\ref{eq:3}) can be expressed in terms of various form factors. For the vector and axial vector form factors, the Bauer-Stech-Wirbel (BWS) forms are more frequently used and defined as~\cite{Wirbel:1985ji},
\begin{eqnarray}\label{eq:5}
\notag
&&\left\langle \chi _{c0}(p') \right|\bar c(0)\gamma _\mu \gamma _5 b(0)\left| B_c(p) \right\rangle = F_0^{{B_c} \to \chi _{c0}}(q^2)\frac{m_{B_c}^2 - m_{\chi _{c0}}^2}{q^2}q_\mu \\
&&+ F_1^{{B_c} \to \chi _{c0}}(q^2)\left( p_\mu + p'_\mu - \frac{m_{B_c}^2 - m_{\chi _{c0}}^2}{q^2}q_\mu  \right) 
\end{eqnarray}
\begin{eqnarray}\label{eq:6}
\notag
&&\left\langle \chi _{c1}[{h_c}](p') \right|\bar c(0)\gamma _\mu b(0)\left| {B_c}(p) \right\rangle  \\
\notag
&&= - i\big(m_{B_c} - m_{\chi _{c1}[h_c]}\big)V_1^{B_c \to \chi _{c1}[h_c]}(q^2)\epsilon ^{\chi_{c1}[h_c]*}_{\mu} \\
\notag
&&+i V_2^{B_c \to \chi _{c1}[h_c]}(q^2)\frac{\epsilon ^{\chi_{c1}[h_c]*} \cdot P}{m_{B_c} - m_{\chi _{c1}[h_c]}}P_\mu \\
\notag
&& + 2im_{\chi _{c1}[h_c]}\left[ V_3^{B_c \to \chi _{c1}[h_c]}(q^2) - V_0^{B_c \to \chi _{c1}[h_c]}(q^2) \right] \frac{\epsilon ^{\chi_{c1}[h_c]*} \cdot P}{q^2}q_\mu \\ 
\notag
&&\left\langle \chi _{c1}[h_c](p') \right|\bar c(0)\gamma _\mu \gamma _5b(0)\left| B_c(p) \right\rangle  \\
&&= - \frac{A^{B_c \to \chi _{c1}[h_c]}(q^2)}{m_{B_c} - m_{\chi _{c1}[h_c]}}\varepsilon_{\mu \alpha \beta \gamma }\epsilon ^{\chi_{c1}[h_c]*}_{\alpha}P_\beta q_\gamma
\end{eqnarray}
\begin{eqnarray}\label{eq:7}
\notag
&&\left\langle \chi _{c2}(p') \right|\bar c(0)\gamma _\mu b(0)\left| B_c(p) \right\rangle  = h^{B_c \to \chi _{c2}}(q^2)\varepsilon_{\mu \beta \rho \sigma }\epsilon^{\chi_{c2}*}_{\beta \lambda }P^\lambda P^\rho q^\sigma\\
\notag
&&\left\langle \chi _{c2}(p') \right|\bar c(0)\gamma_\mu \gamma _5b(0)\left| B_c(p) \right\rangle = -i\ k^{B_c \to \chi _{c2}}(q^2)\epsilon^{\chi_{c2}*}_{\mu \beta }P^\beta \\
&&-i \epsilon^{\chi_{c2}*}_{\beta \lambda}P^\beta P^\lambda \left[ b_ + ^{B_c \to \chi _{c2}}(q^2)P_\mu + b_-^{B_c \to \chi _{c2}}(q^2)q_\mu \right]  
\end{eqnarray}
where $P=p+p'$ and $q=p-p'$. $F_{0}$ and $F_{1}$ are axial vector form factors of $\chi_{c0}$, the vector form factors of $\chi_{c0}$ do not exist due to the conservation of angular momentum. $A$ and $V_{0}$/$V_{1}$/$V_{2}$ are axial vector and vector form factors of $\chi_{c1}$ and $h_{c}$. $V_{3}$ in Eq. (\ref{eq:6}) has the following form,
\begin{eqnarray}\label{eq:8}
\notag
V_{3}^{B_c \to \chi_{c1}[h_c]}(q^2)&&=\frac{m_{B_c}-m_{\chi_{c1}[h_{c}]}}{2m_{\chi_{c1}[h_{c}]}}V_{1}^{B_c \to \chi_{c1}[h_c]}(q^2)\\
&&-\frac{m_{B_c}+m_{\chi_{c1}[h_{c}]}}{2m_{\chi_{c1}[h_{c}]}}V_{2}^{B_c \to \chi_{c1}[h_c]}(q^2)
\end{eqnarray}
$h$ and $k/b_{+}/b_{-}$ in Eq. (\ref{eq:7}) are vector and axial vector form factors of $B_{c}\to\chi_{c2}$. In order to simplify the calculations of scatter amplitude, the following substitutions are commonly used~\cite{Wang:2009mi},
\begin{eqnarray}\label{eq:9}
\notag
A^{B_c \to \chi_{c2}}(q^2)&&=-\big(m_{B_c}-m_{\chi_{c2}}\big)h^{B_c \to \chi_{c2}}(q^2) \\
\notag
V_{1}^{B_c \to \chi_{c2}}(q^2)&&=-\frac{k^{B_c \to \chi_{c2}}(q^2)}{m_{B_c}-m_{\chi_{c2}}} \\
\notag
V_{2}^{B_c \to \chi_{c2}}(q^2)&&=\big(m_{B_c}-m_{\chi_{c2}}\big)b_{+}^{B_c \to \chi_{c2}}(q^2) \\
\notag
V_{0}^{B_c \to \chi_{c2}}(q^2)&&=\frac{m_{B_c}-m_{\chi_{c2}}}{2m_{\chi_{c2}}}V_{1}^{B_c \to \chi_{c2}}(q^2) \\
\notag
&&-\frac{m_{B_c}+m_{\chi_{c2}}}{2m_{\chi_{c2}}}V_{2}^{B_c \to \chi_{c2}}(q^2)-\frac{q^2}{2m_{\chi_{c2}}}b_{-}^{B_c \to \chi_{c2}}(q^2) \\
\end{eqnarray}
The tensor form factors of $B_{c}\to\chi_{cJ}$ and $h_{c}$ are defined as follows~\cite{Colangelo:2022awx}, 
\begin{eqnarray}\label{eq:10}
\left\langle \chi _{c0}(p') \right|\bar c(0)\sigma _{\mu \nu }b(0)\left| B_c(p) \right\rangle =  - \frac{T^{B_c \to \chi _{c0}}(q^2)}{m_{B_c} - m_{\chi _{c0}}}\varepsilon _{\mu \nu \alpha \beta }p^\alpha p'^\beta 
\end{eqnarray}
\begin{eqnarray}\label{eq:11}
\notag
&&\left\langle \chi _{c1}[h_c](p') \right|\bar c(0)\sigma _{\mu \nu }\gamma _5b(0)\left| B_c(p) \right\rangle  \\
\notag
&&= - T_0^{B_c \to \chi _{c1}[h_c]}(q^2)\frac{\epsilon ^{\chi_{c1}[h_c]*} \cdot q}{\big(m_{B_c} - {m_{\chi _{c1}[h_c]}}\big)^2}\varepsilon _{\mu \nu \alpha \beta }p^\alpha p'^\beta  \\
\notag
&&+ T_1^{B_c \to \chi _{c1}[h_c]}(q^2)\varepsilon _{\mu \nu \alpha \beta }p_\alpha \epsilon^{\chi_{c1}[h_c]*}_{\beta}\\
 &&+ T_2^{B_c \to \chi _{c1}[h_c]}(q^2)\varepsilon _{\mu \nu \alpha \beta}p'_\alpha \epsilon^{\chi_{c1}[h_c]*}_{\beta}
\end{eqnarray}
\begin{eqnarray}\label{eq:12}
\notag
&&\left\langle \chi _{c2}(p') \right|\bar c(0)\sigma _{\mu \nu }b(0)\left| B_c(p) \right\rangle \\
\notag
&& = \frac{q^\tau}{m_{B_c}}\left[ T_0^{B_c \to \chi _{c2}}(q^2)\frac{\epsilon^{\chi_{c2}*}_{\rho \tau}q^\rho }{\big(m_{B_c} - m_{\chi _{c2}}\big)^2}\varepsilon _{\mu \nu \chi \delta }p^\chi p'^\delta \right.\\
\notag
&&\left. + T_1^{B_c \to \chi _{c2}}(q^2)\varepsilon _{\mu \nu \chi \delta}\epsilon ^{\chi_{c2}*}_{\chi \tau}p_\delta + T_2^{B_c \to \chi _{c2}}(q^2)\varepsilon _{\mu \nu \chi \delta}\epsilon^{\chi_{c2}*}_{\chi \tau }p'_\delta \right] \\
\end{eqnarray}
where $T$ and $T_{0}/T_{1}/T_{2}$ are tensor form factors of $\chi_{c0}$ and $\chi_{c1}/\chi_{c2}/h_{c}$. 
 
From these above equations, the correlation functions in phenomenological side can be obtained and be expanded into different tensor structures. In principle, each structure can be used to carry out the calculations of form factors and will lead to same results. However, the final results obtained by different structures have different uncertainties because of the truncation of operator product expansion (OPE) in QCD side and different contributions of higher resonances and continuum states~\cite{Bracco:1999xe}. In this article, we will adopt the traditional way to choose the appropriate structure, where the pole contributions should be large than 50\% and the stability of Borel window should be satisfied. The details will discussed in next sections.

\subsection{The QCD side}\label{sec2.2}
In QCD side, we perform the operator product expansion (OPE) of correlation function by contracting all of the quark fields operator with Wick's theorem. After this step, the correlation functions of all processes in QCD side are expressed as follows, 
\begin{eqnarray}\label{eq:13}
\notag
\Pi _\mu ^{\chi _{c0}QCD}(p,p') &&=  - i\int d^4xd^4ye^{ip'x}e^{i(p - p')y} \\
\notag
&& \times Tr\big[C^{km}(-x)C^{mn}(x - y)\gamma _\mu \gamma _5B^{nk}(y)\gamma _5\big]\\
\notag
\Pi _{\mu \nu }^{\chi _{c0}QCD}(p,p') &&=  - i\int d^4xd^4ye^{ip'x}e^{i(p - p')y} \\
&& \times Tr\big[C^{km}(-x)C^{mn}(x - y)\sigma _{\mu \nu }B^{nk}(y)\gamma _5\big]
\end{eqnarray}
\begin{eqnarray}\label{eq:14}
\notag
\Pi _{1\mu \nu }^{\chi _{c1}QCD}(p,p') &&=  - i\int d^4xd^4ye^{ip'x}e^{i(p - p')y} \\
\notag
&& \times Tr[C^{km}(-x)\gamma _\mu \gamma _5C^{mn}(x - y)\gamma _\nu B^{nk}(y)\gamma _5]\\
\notag
\Pi _{2\mu \nu }^{\chi _{c1}QCD}(p,p') &&=  - i\int d^4xd^4ye^{ip'x}e^{i(p - p')y} \\
\notag
&& \times Tr\big[C^{km}(-x)\gamma _\mu \gamma _5C^{mn}(x - y)\gamma _\nu \gamma _5B^{nk}(y)\gamma _5\big]\\
\notag
\Pi _{\mu \nu \sigma }^{\chi _{c1}QCD}(p,p') &&=  - i\int d^4xd^4ye^{ip'x}e^{i(p - p')y} \\
\notag
&& \times Tr\big[C^{km}(-x)\gamma _\mu \gamma _5C^{mn}(x - y)\sigma _{\nu \sigma }\gamma _5B^{nk}(y)\gamma _5\big] \\
\end{eqnarray}
\begin{eqnarray}\label{eq:15}
\notag
\Pi _{1\mu \nu \sigma }^{h_cQCD}(p,p') &&=  - i\int d^4xd^4ye^{ip'x}e^{i(p - p')y} \\
\notag
&& \times Tr\big[C^{km}(-x)\sigma _{\mu \nu }C^{mn}(x - y)\gamma _\sigma B^{nk}(y)\gamma _5\big]\\
\notag
\Pi _{2\mu \nu \sigma }^{h_cQCD}(p,p') &&=  - i\int d^4xd^4ye^{ip'x}e^{i(p - p')y} \\
\notag
&& \times Tr\big[C^{km}(-x)\sigma _{\mu \nu }C^{mn}(x - y)\gamma _\sigma \gamma _5B^{nk}(y)\gamma _5\big]\\
\notag
\Pi _{\mu \nu \alpha \beta }^{h_cQCD}(p,p') &&=  - i\int d^4xd^4ye^{ip'x}e^{i(p - p')y} \\
\notag
&& \times Tr\big[C^{km}(-x)\sigma _{\mu \nu }C^{mn}(x - y)\sigma _{\alpha \beta }\gamma _5B^{nk}(y)\gamma _5\big] \\
\end{eqnarray}
\begin{eqnarray}\label{eq:16}
\notag
\Pi _{1\mu \nu \sigma }^{\chi _{c2}QCD}(p,p') &&= \int d^4xd^4ye^{ip'x}e^{i(p - p')y} \\
\notag
&&\times Tr\big[C^{km}(-x)\Gamma _{\mu \nu }C^{mn}(x - y)\gamma _\sigma B^{nk}(y)\gamma _5\big]\\
\notag
\Pi _{2\mu \nu \sigma }^{\chi _{c2}QCD}(p,p') &&= \int d^4xd^4ye^{ip'x}e^{i(p - p')y} \\
\notag
&&\times Tr\big[C^{km}(-x)\Gamma _{\mu \nu }C^{mn}(x - y)\gamma _\sigma\gamma _5B^{nk}(y)\gamma _5\big]\\
\notag
\Pi _{\mu \nu \alpha \beta }^{\chi _{c2}QCD}(p,p') &&= \int d^4xd^4ye^{ip'x}e^{i(p - p')y} \\
\notag
&& \times Tr\big[C^{km}(-x)\Gamma _{\mu \nu }C^{mn}(x - y)\sigma _{\alpha \beta }B^{nk}(y)\gamma _5\big] \\
\end{eqnarray}
where $\Gamma_{\mu\nu}=\frac{1}{2}(\gamma_{\mu} \mathord{\buildrel{\lower3pt\hbox{$\scriptscriptstyle\leftrightarrow$}} 
	\over D}_{\nu}+\gamma_{\nu}\mathord{\buildrel{\lower3pt\hbox{$\scriptscriptstyle\leftrightarrow$}} 
	\over D}_{\mu}-\frac{1}{2}g_{\mu\nu}{\mathord{\buildrel{\lower3pt\hbox{$\scriptscriptstyle\leftrightarrow$}} 
		\over{\slashed D}}})$, $C^{ij}(x)$ and $B^{ij}(x)$ are the full propagator of $c$ and $b$ quarks, and they can uniformly be represented as the following forms~\cite{Wang:2014yza},
\begin{eqnarray}\label{eq:17}
\notag
Q^{ij}(x) &&= \frac{i}{(2\pi )^4}\int d^4 k e^{ - ik \cdot x} \left \{\frac{\delta ^{ij}}{\slashed k - {m_Q}} \right. \\
\notag
&& - \frac{g_sG_{\alpha \beta }^nt_{ij}^n}{4}\frac{\sigma ^{\alpha \beta }(\slashed{k} + {m_Q}) + (\slashed{k} + {m_Q})\sigma ^{\alpha \beta }}{(k^2 - m_Q^2)^2}\\
\notag
&& \left. - \frac{g_s^2({t^a}{t^b})_{ij}G_{\alpha \beta }^aG_{\mu \nu }^b(f^{\alpha \beta \mu \nu } + f^{\alpha \mu \beta \nu } + f^{\alpha \mu \nu \beta })}{4(k^2 - m_Q^2)^5} + ... \right\} \\
\end{eqnarray}
Here, $Q$ denotes $C$ or $B$, $t^{n}=\frac{\lambda^{n}}{2}$, $\lambda^{n}(n=1,...,8)$ are the Gell-Mann matrices, $i$ and $j$ are color indices, $\sigma_{\alpha\beta}=\frac{i}{2}[\gamma_{\alpha},\gamma_{\beta}]$, and $f^{\alpha\beta\mu\nu}$ have the following form,
\begin{eqnarray}\label{eq:18}
\notag
f^{\alpha \beta \mu \nu }&&= (\slashed k + m_Q)\gamma ^\alpha (\slashed k + m_Q)\gamma ^\beta (\slashed k + m_Q)\\
&& \times \gamma ^\mu (\slashed k + m_Q)\gamma ^\nu (\slashed k + m_Q)
\end{eqnarray}
Substituting with the full propagator in Eqs. (\ref{eq:13})$\sim$(\ref{eq:16}), the correlation functions in QCD side can also be  expanded in different tensor structures, 
\begin{eqnarray}\label{eq:19}
\notag
\Pi _\mu ^{\chi _{c0}QCD}(p,p') &&= \Pi _1^{\chi _{c0}QCD}p_\mu  + \Pi _2^{\chi _{c0}QCD}p'_\mu \\
\Pi _{\mu \nu }^{\chi _{c0}QCD}(p,p') &&= \Pi _3^{\chi _{c0}QCD}\varepsilon _{\mu \nu \alpha \beta}p^\alpha p'^\beta 
\end{eqnarray}
\begin{eqnarray}\label{eq:20}
\notag
\tilde \Pi _{1\lambda \nu }^{\chi _{c1}QCD}(p,p') &&=\big(g_{\lambda\mu}-\frac{p'_\lambda p'_\mu}{p'^2}\big)\Pi_{1\mu\nu}^{\chi _{c1}QCD}(p,p')        \\
\notag
&&= \Pi _1^{\chi _{c1}QCD}g_{\lambda \nu } + \Pi _2^{\chi _{c1}QCD}p_\lambda p_\nu \\
\notag
&&+ \Pi _3^{\chi _{c1}QCD}p_\lambda p'_\nu + ...\\
\notag
\tilde \Pi _{2\lambda \nu }^{\chi _{c1}QCD}(p,p') &&=\big(g_{\lambda\mu}-\frac{p'_\lambda p'_\mu}{p'^2}\big)\Pi _{2\mu\nu }^{\chi _{c1}QCD}(p,p')\\
\notag
&&= \Pi _4^{\chi _{c1}QCD}\varepsilon _{\lambda \nu \alpha \beta }p^\alpha p'^\beta \\
\notag
\tilde \Pi _{\lambda \nu \sigma }^{\chi _{c1}QCD}(p,p') &&=\big(g_{\lambda\mu}-\frac{p'_\lambda p'_\mu}{p'^2}\big)\Pi _{\mu \nu \sigma }^{\chi _{c1}QCD}(p,p')     \\
\notag
&&= \Pi _5^{\chi _{c1}QCD}\varepsilon _{\lambda \nu \sigma \alpha }p^\alpha + \Pi _6^{\chi _{c1}QCD}\varepsilon _{\lambda \nu \sigma \alpha }p'^\alpha \\
&& + \Pi _7^{\chi _{c1}QCD}p_\lambda \varepsilon _{\nu \sigma \alpha \beta }p^\alpha p'^\beta  + ...
\end{eqnarray}
\begin{eqnarray}\label{eq:21}
\notag
\tilde \Pi _{1\lambda \chi \sigma }^{h_cQCD}(p,p') &&=\big(g_{\lambda\mu}-\frac{p'_\lambda p'_\mu}{p'^2}\big)\big(g_{\chi\nu}-\frac{p'_\chi p'_\nu}{p'^2}\big) \Pi _{1\mu \nu \sigma }^{h_cQCD}(p,p')\\
\notag
&&= \Pi _1^{h_cQCD}\varepsilon _{\lambda \sigma \chi \alpha }p^\alpha + \Pi _2^{h_cQCD}p_\sigma \varepsilon _{\lambda \chi \alpha \beta }p^\alpha p'^\beta \\
\notag
&& + \Pi _3^{h_cQCD}p'_\sigma \varepsilon _{\lambda \chi \alpha \beta }p^\alpha p'^\beta  + ...\\
\notag
\tilde \Pi _{2\lambda \chi \sigma }^{h_cQCD}(p,p') &&=\big(g_{\lambda\mu}-\frac{p'_\lambda p'_\mu}{p'^2}\big)\big(g_{\chi\nu}-\frac{p'_\chi p'_\nu}{p'^2}\big) \Pi _{2\mu \nu \sigma }^{h_cQCD}(p,p')   \\
\notag
&&= \Pi _4^{h_cQCD}p_\lambda g_{\sigma \chi } + ...\\
\notag
\bar \Pi _{\alpha \chi }^{h_cQCD}(p,p') &&=g^{\lambda\beta}\big(g_{\lambda\mu}-\frac{p'_\lambda p'_\mu}{p'^2}\big)\big(g_{\chi\nu}-\frac{p'_\chi p'_\nu}{p'^2}\big)\Pi _{\mu \nu \alpha \beta}^{h_cQCD}(p,p')    \\
\notag
&&= \Pi _5^{h_cQCD}g_{\alpha \chi } + \Pi _6^{h_cQCD}p_\alpha p_\chi \\
&& + \Pi _7^{h_cQCD}p'_\alpha p_\chi + ...
\end{eqnarray}
\begin{eqnarray}\label{eq:22}
\notag
\Pi _{1\mu \nu \sigma }^{\chi _{c2}QCD}(p,p') &&= \Pi _1^{\chi _{c2}QCD}p_\mu \varepsilon _{\sigma \nu \alpha \beta }p^\alpha p'^\beta \\
\notag
\Pi _{2\mu \nu \sigma }^{\chi _{c2}QCD}(p,p') &&= \Pi _2^{\chi _{c2}QCD}p_\mu g_{\nu \sigma } + \Pi _3^{\chi _{c2}QCD}p_\mu p_\nu p_\sigma \\
\notag
&& + \Pi _4^{\chi _{c2}QCD}p'_\mu p'_\nu p'_\sigma \\
\notag
\Pi _{\mu \nu \alpha \beta }^{\chi _{c2}QCD}(p,p') &&= \Pi _5^{\chi _{c2}QCD}g_{\mu \nu }\varepsilon _{\alpha \beta \chi \lambda }p^\chi p'^\lambda + \Pi _6^{\chi _{c2}QCD}p_\mu \varepsilon _{\nu \alpha \beta \lambda }p^\lambda \\
&& + \Pi _7^{\chi _{c2}QCD}p_\mu \varepsilon _{\nu \alpha \beta \lambda }p'^\lambda  + ...
\end{eqnarray}
The projection operator $\big(g_{\lambda\mu}-\frac{p'_{\lambda}p'_{\mu}}{p'^{2}}\big)$ in Eq. (\ref{eq:20}) is introduced to eliminate the coupling of axial vector current $\bar{c}\gamma_{\mu}\gamma_{5}c$ with pseudoscalar charmonium $\eta_{c}$, and $\big(g_{\lambda\mu}-\frac{p'_{\lambda}p'_{\mu}}{p'^{2}}\big)\big(g_{\chi\nu}-\frac{p'_{\chi}p'_{\nu}}{p'^{2}}\big)$ in Eq. (\ref{eq:21}) is to eliminate the coupling of tensor current $\bar{c}\sigma_{\mu\nu}c$ with vector charmonium $J/\psi$. $\Pi^{QCD}_{i}$ in the right side of above equations are named as the scalar invariant amplitude, and can  be divided into perturbative and non-perturbative parts,
\begin{eqnarray}\label{eq:23}
\Pi _i^{QCD}(p,p') &&= \Pi^{pert}_{i}(p,p')+\Pi^{non-pert}_{i}(p,p')
\end{eqnarray}
where the non-perturbative part include the two gluons condensate $\langle g_{s}^{2}GG\rangle$, three gluons condensate $g_{s}^{3}\langle fGGG\rangle$, and higher dimensions condensate terms. In our previous work~\cite{Wu:2024gcq,Lu:2024tgy}, the contribution of three gluon condensate is small, about one-tenth that of two-gluon condensate. The higher dimension terms such as the four gluon condensate are very small and further suppressed by $O(\alpha_{s}^{2})$ $(\alpha_{s}=g_{s}^{2}/4\pi)$, so can be safely neglected in our calculations. Thus, we only reserve the two gluon condensate term as the non-perturbative contribution in our calculations. The scalar invariant amplitude can be written as the following form according to the double dispersion relation,
\begin{eqnarray}\label{eq:24}
\Pi _i^{QCD}(p,p') &&= \int\limits_{s_1}^{\infty} ds\int\limits_{u_1}^{\infty} du  \frac{\rho _i^{QCD}(s,u,q^2)}{(s - p^2)(u - p'^2)}
\end{eqnarray}
where $\rho^{QCD}(s,u,q^2)$ is the QCD spectral density and can be expressed as,
\begin{eqnarray}\label{eq:25}
\rho _i^{QCD}(s,u,q^2) &&= \rho ^{pert}(s,u,q^2) + \rho ^{\left\langle g_s^2GG \right\rangle }(s,u,q^2)
\end{eqnarray}
with $s=p^2$, $u=p'^2$ and $q=p-p'$. $s_1$ and $u_1$ in Eq. (\ref{eq:24}) are the kinematic limits for $B_{c}$ meson and $P$-wave charmonium, and they are taken as $(m_{b}+m_{c})^{2}$ and $4m_{c}^{2}$, respectively.

\begin{figure}
	\centering
	\includegraphics[width=8.5cm]{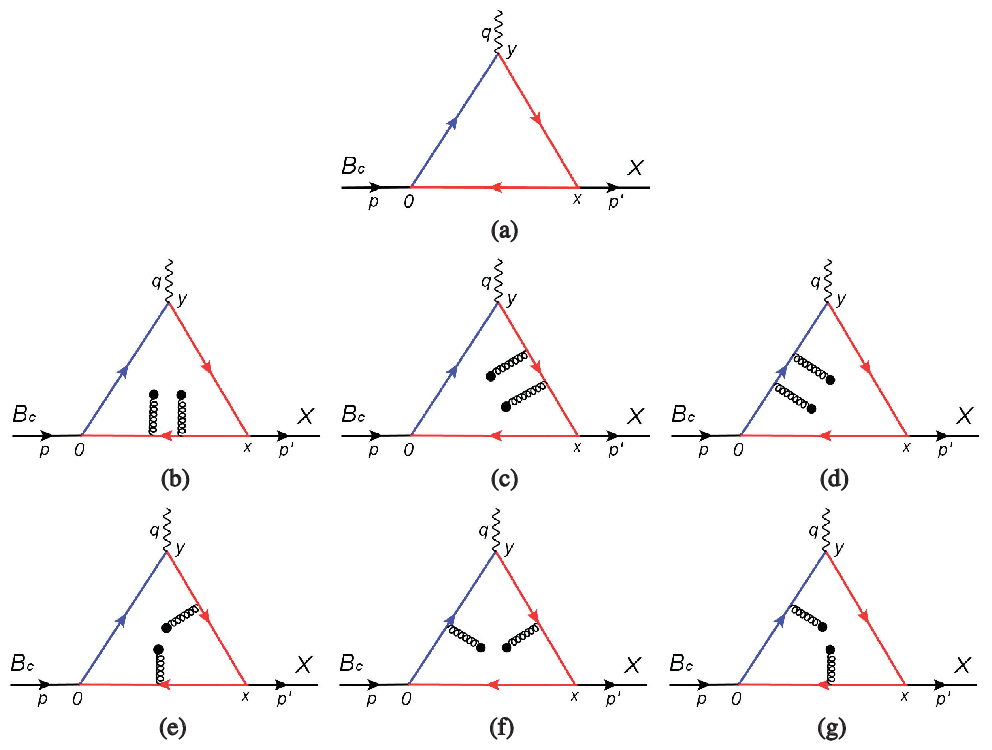}
	\caption{The Feynman diagrams for the perturbative part (a) and gluon condensate (b)$\sim$(g), where the blue and red solid lines denote the $b$ and $c$ quark lines, respectively. The black loop lines are the gluon lines.}
	\label{Feynman diagrams}
\end{figure}

The QCD spectral density of the perturbative part and two gluon condensate can be obtained by using the Cutkoskys's rules~\cite{Wang:2007ys,Shi:2019hbf}, the calculation details can be found in Ref. ~\cite{Wu:2024gcq}. The corresponding Feynman diagrams are shown in Fig. \ref{Feynman diagrams}.
We take the change of variables $p^{2}\to- P^{2}$, $p'^{2}\to-P'^{2}$ and $q^{2}\to-Q^{2}$ and perform double Borel transforms~\cite{Ioffe:1982qb} for the variables $P^{2}$ and $P'^{2}$ to both phenomenological and QCD sides. The variables $P^{2}$ and $P'^{2}$ will be replaced by Borel parameters $T_{1}^{2}$ and $T_{2}^{2}$, respectively. For simplicity, the relations $T^{2}=T_{1}^{2}$ and $T_{2}^{2}=kT_{1}^{2}=kT^{2}$ are usually used to reduce the two Borel parameters, where $k=m_{X}^{2}/m_{B_{c}}^{2}$~\cite{Bracco:2011pg}, $X$ denote the P-wave charmonium. The plausibility for this reduction will be discussed in the next section. After matching the phenomenological and QCD sides by using quark-hadron duality condition, the QCD sum rules for the form factors will be obtained. Taking the axial vector form factors for $B_{c}\to \chi_{c0}$ as an example, these form factors can be obtained by combining two invariant amplitudes $\Pi^{\chi_{c0}QCD}_{1}$ and $\Pi^{\chi_{c0}QCD}_{2}$ in Eq. (\ref{eq:19}), and have the following forms,
\begin{eqnarray}\label{eq:26}
\notag
F_0(Q^2) &&=  - \frac{1}{2\mathcal{X}\mathcal{Y}}e^{\frac{m_{B_{c}}^2}{T_1^2}+\frac{m_{X}^2}{T_2^2}} \\
\notag
&&\times\int\limits_{s_1}^{s_0}\int\limits_{u_1}^{u_0} dsdu \big[\big(\mathcal{X} + 1)\rho _1^{{\chi _{c0}}QCD}(s,u,{Q^2}\big) \\
\notag
&& + \big(\mathcal{X} - 1\big)\rho _2^{\chi _{c0}QCD}(s,u,Q^2)\big]e^{- \frac{s}{T_1^2}-\frac{u}{T_2^2}} \\
\notag
F_1(Q^2) &&=  -\frac{1}{2\mathcal{Y}}e^{\frac{m_{B_{c}}^2}{T_1^2}+\frac{m_{X}^2}{T_2^2}} \\
\notag
&&\times\int\limits_{s_1}^{s_0}\int\limits_{u_1}^{u_0} dsdu\big[\rho _1^{\chi _{c0}QCD}(s,u,Q^2)\\
&& + \rho _2^{\chi _{c0}QCD}(s,u,Q^2)\big]e^{-\frac{s}{T_1^2}-\frac{u}{T_2^2}}
\end{eqnarray}
where, 
\begin{eqnarray}\label{eq:27}
\notag
\mathcal{X} &&=  - \frac{m_{B_c}^2 - m_{\chi _{c0}}^2}{Q^2}\\
\mathcal{Y} &&= \frac{f_{\chi _{c0}}f_{B_c}m_{\chi _{c0}}m_{B_c}^2}{m_b + m_c} 
\end{eqnarray}
In Eq. (\ref{eq:26}), $s_{0}$ and $u_{0}$ are the threshold parameters which are introduced to eliminate the contributions of higher resonances and continuum states. They usually fulfill the relations $s_{0}=(m_{B_{c}}+\delta)^{2}$ and $u_{0}=(m_{X}+\delta)^{2}$, $\delta$ is the mass gap between the ground and first excited states and is commonly taken as the value of $0.4\sim0.6$ GeV~\cite{Bracco:2011pg}.

\begin{figure}
	\centering
	\includegraphics[width=5cm, trim=0 120 0 50, clip]{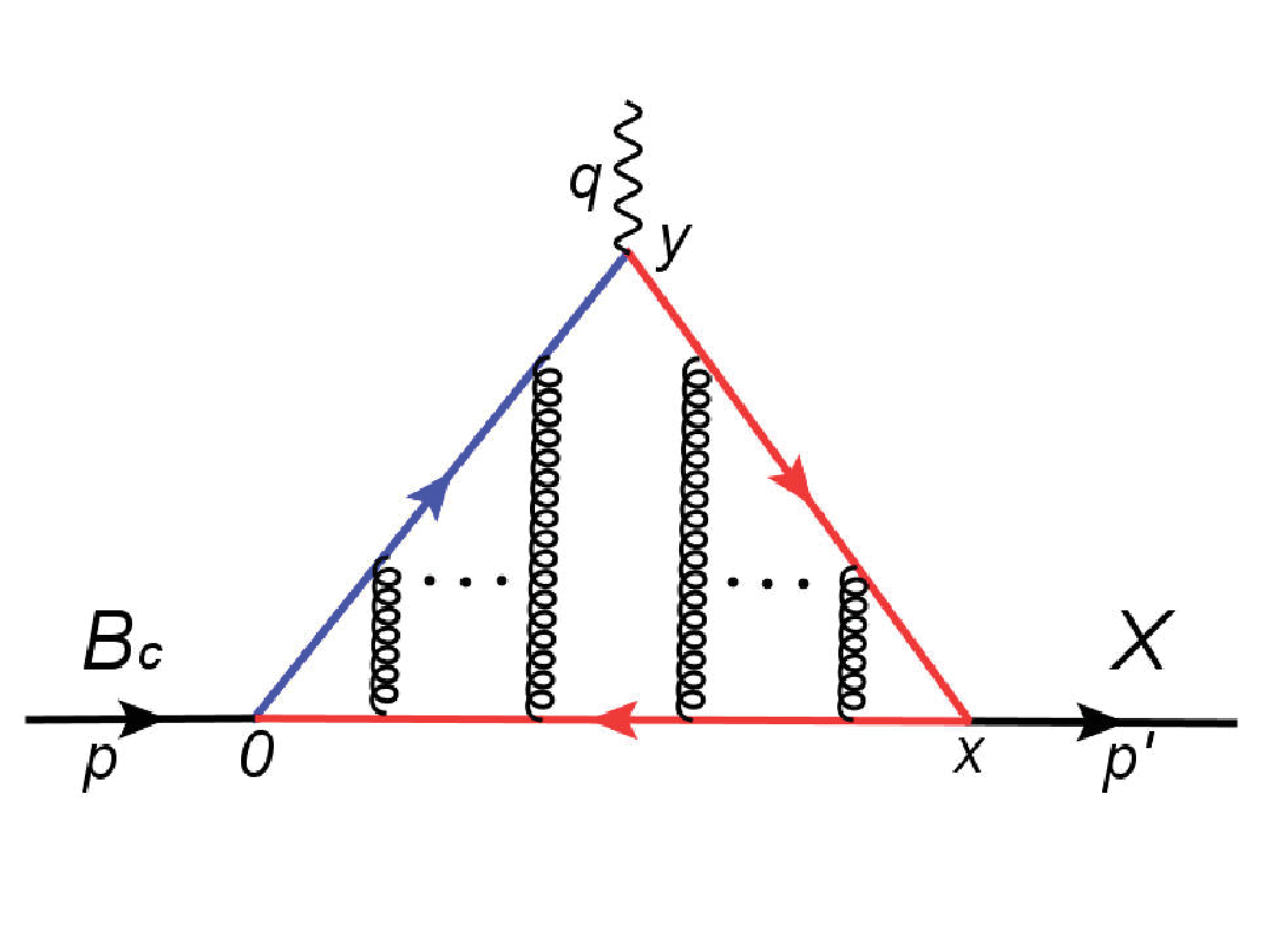}
	\caption{The Feynman diagram for the Coulomb-like interaction.}
	\label{correction}
\end{figure}

For the full heavy quark system, the quark condensate does not exist, major non-perturabative contributions are gluon condensates. In addition, the next-to-leading order correction for the perturbative contribution is also important for the accuracy of final results. However, the rigorous calculation for the loop diagrams in three-point QCD sum rules is difficult. The Coulomb-like $\alpha_{s}/v$ correlation is proposed to ameliorate this problem~\cite{Kiselev:2000pp}, and it can be illustrated in Fig. \ref{correction}. In non-relativistic approximation, this correction will contribute a renormalization coefficient for the spectral density of the perturbative part, 
\begin{eqnarray}\label{eq:28}
\notag
\rho _c^{pert}(s,u,Q^2) &&= \sqrt {\frac{4\pi \alpha _s^C}{3v_1}{\left[1 - \exp \big( - \frac{4\pi \alpha _s^C}{3v_1}\big) \right]^{ - 1}}} \\
\notag
&& \times \sqrt {\frac{4\pi \alpha _s^C}{3v_2}{\left[1 - \exp \big( - \frac{4\pi \alpha _s^C}{3v_2}\big) \right]^{ - 1}}}\rho ^{pert}(s,u,Q^2) \\
\end{eqnarray}
where $\alpha_s^C\approx \alpha_s(\mu)$, $v_{1}$ and $v_{2}$ are the relative velocities of heavy quarks in the $B_{c}$ meson and $P$-wave charmonia, and can be expressed as,
\begin{eqnarray}\label{eq:29}
\notag
v_1 &&= \sqrt {1 - \frac{4m_bm_c}{s - {(m_b- m_c)^2}}} \\
v_2 &&= \sqrt {1 - \frac{4m_c^2}{u}} 
\end{eqnarray}
In this study, we take into account the contribution of this correction and briefly discuss its impact on the results.

\section{Weak decay of $B_{c}$ to P-Wave charmonia}\label{sec3}
This section is devoted to analyzing the semileptonic and nonleptonic decay processes of $B_{c}$ meson by using the vector and axial vector form factors obtained in previous section.
\subsection{Semileptonic decay}
The semileptonic decay $B_{c}\to Xl\bar{\nu_{l}}$ $(l=e,\mu$ and $\tau)$ can be described by the following effective Hamiltonian,
\begin{eqnarray}\label{eq:30}
	H_{eff} = \frac{G_F}{\sqrt 2}\sum\limits_{l = e,\mu ,\tau } {V_{cb}\bar c\gamma _\mu (1 - \gamma _5)b\bar \nu _l\gamma _\mu(1 - \gamma _5)l} 
\end{eqnarray}
where $G_{F}=1.16637\times 10^{-5}$ GeV$^{-2}$ is the Fermi constant, $V_{cb}$ is CKM matrix element. The transition matrix element can be written as,
\begin{eqnarray}\label{eq:31}
	\notag
	T &&= \left\langle X(p')l(l)\bar {v}_l(v) \right|H_{eff}\left| B_c(p) \right\rangle \\
	\notag
	&& = \frac{G_F}{\sqrt 2}V_{cb}\left\langle X(p') \right|\bar c\gamma _\mu (1 - \gamma _5)b\left| B_c(p) \right\rangle \\
	&& \times \left\langle l(l)\bar {v}_l(v) \right|\bar \nu _l\gamma _\mu (1 - \gamma _5)l\left| 0 \right\rangle 
\end{eqnarray}
In the above equation, the $B_{c}\to X$ transition matrix element can be represented as the vector and axial vector form factors. The leptonic matrix element can be formulated as the following form by using perturbative field theory approach,
\begin{eqnarray}\label{eq:32}
	\left\langle l(l)\bar {v}_l(v) \right|\bar \nu _l\gamma _\mu (1 - \gamma _5)l\left| 0 \right\rangle  = {\bar u_{v,s}}\gamma _\mu (1 - \gamma _5)u_{ - l,s'}
\end{eqnarray} 
where $u_{v,s}$ and $u_{-l,s'}$ are the spinor wave functions of $\bar{\nu_{l}}$ and $l$, the subscript $v(l)$ and $s(s')$ denote the four momentum and spin of corresponding particles.
\subsection{Nonleptonic decay}
The nonleptonic decays $B_{c}\to X P$ and $B_{c}\to X V$ where $X$ represent the $P$-wave charmonia $\chi_{cJ}$ $(J=0,1,2)$ and $h_{c}$, $P$ denote the light pseudoscalar meson $\pi$ or $K$ and $V$ denote the light vector meson $\rho$ or $K^{*}$ can be uniformly described by the following effective Hamiltonian, 
\begin{eqnarray}\label{eq:33}
H_{eff} = \frac{G_F}{\sqrt 2}\sum\limits_{q = d,s} {V_{cb}V_{uq}^*a_1\bar c\gamma _\mu (1 - \gamma _5)b\bar q\gamma _\mu (1 - \gamma _5)u} 
\end{eqnarray}
where $V_{cb}$ and $V_{uq}$ are CKM matrix elements and $a_{1}$ is the Wilson coefficient. The transition matrix for these decay processes can be expressed as the following form by using the naive factorization approach (NFA)~\cite{Bauer:1984zv,Bauer:1986bm},
\begin{eqnarray}\label{eq:34}
\notag
T &&= \left\langle X(p')P[V](q) \right|H_{eff}\left| B_c(p) \right\rangle \\
\notag
&& = \frac{G_F}{\sqrt 2}V_{cb}V_{uq}^*a_1\left\langle X(p') \right|\bar c\gamma _\mu (1 - \gamma _5)b\left| B_c(p) \right\rangle \\
&& \times \left\langle P[V](q) \right|\bar q\gamma _\mu (1 - \gamma _5)u\left| 0 \right\rangle 
\end{eqnarray}
The meson vacuum matrix elements can be defined as the following forms,
\begin{eqnarray}\label{eq:35}
\notag
\left\langle P(q) \right|\bar q\gamma _\mu (1 - \gamma _5)u\left| 0 \right\rangle  &&=-if_Pq_\mu \\
\left\langle V(q) \right|\bar q\gamma _\mu (1 - \gamma _5)u\left| 0 \right\rangle  &&= f_Vm_V\epsilon _\mu ^* 
\end{eqnarray}
where $f_{P}$ and $f_{V}$ are the decay constants of pseudoscalar and vector meson, $\epsilon_{\mu}$ is the polarization vector of vector meson.

\section{Numerical results and discussions}\label{sec4}
\subsection{Numerical results of form factors}
As important input parameters, the heavy quark masses are energy scale dependent, which can be expressed as the following renormalization group equation (RGE),
\begin{eqnarray}\label{eq:36}
\notag
m_Q(\mu ) &&= m_Q(m_Q)\left[ \frac{\alpha _s(\mu )}{\alpha _s(m_Q)} \right]^{\frac{12}{33 - 2N_f}}\\
\notag
\alpha _s(\mu ) &&= \frac{1}{b_0 t}\left[1 - \frac{b_1}{b_0^2}\frac{\log t}{t} \right.\\
&&\left. + \frac{b_1^2({\log^2}t - \log t - 1) + b_0b_2}{b_0^4t^2} \right]
\end{eqnarray}
where $t=\log(\frac{\mu^{2}}{\Lambda_{QCD}^{2}})$, $b_{0}=\frac{33-2N_{f}}{12\pi}$, $b_{1}=\frac{153-19N_{f}}{24\pi^{2}}$ and $b_{2}=\frac{2857-\frac{5033}{9}N_{f}+\frac{325}{27}N_{f}^{2}}{128\pi^{3}}$. $\Lambda_{QCD}=213$ MeV for the flavors $N_{f}=5$ in this work~\cite{ParticleDataGroup:2024cfk}. The $\mathrm{\overline{MS}}$ masses of $c$ and $b$ quarks are taken from the Particle Date Group (PDG)~\cite{ParticleDataGroup:2024cfk}, where $m_{c}(m_{c})=1.275\pm0.025$ GeV and  $m_{b}(m_{b})=4.18\pm0.03$ GeV. Based on our previous work~\cite{Wang:2024fwc}, the energy scale $\mu=2$ GeV works well in $B_{c}$ meson system. Thus, this value is still employed in the present work. The energy scale dependence of the form factors will be shown below. The other parameters like masses, decay constants of $B_{c}$ meson and P-wave charmonia and the standard value of vacuum condensate parameter are all listed in Table \ref{IP}.
\begin{table}[htbp]
\begin{ruledtabular}\caption{Input parameters (IP) used to calculate the form factors.}
\label{IP}
\begin{tabular}{c c c c }
IP&Values (GeV)&IP&Values \\ \hline
$m_{B_{c}}$&6.274~\cite{ParticleDataGroup:2024cfk}&$f_{\chi_{c0}}$&0.343 GeV~\cite{Novikov:1977dq}\\
$m_{\chi_{c0}}$&3.414~\cite{ParticleDataGroup:2024cfk}&$f_{\chi_{c1}}$&0.338 GeV~\cite{Novikov:1977dq} \\
$m_{\chi_{c1}}$&3.511~\cite{ParticleDataGroup:2024cfk}&$f_{h_{c}}$&0.235 GeV~\cite{Becirevic:2013bsa} \\
$m_{h_{c}}$&3.525~\cite{ParticleDataGroup:2024cfk}&$f_{\chi_{c2}}$&$0.0111\pm0.0062$ ~\cite{Aliev:2010ac} \\
$m_{\chi_{c2}}$&3.556~\cite{ParticleDataGroup:2024cfk}&$\langle g_{s}^{2}GG\rangle$&$0.88\pm0.15$ GeV$^{4}$~\cite{Narison:2010cg,Narison:2011xe,Narison:2011rn} \\
$f_{B_{c}}$&$0.371\pm0.037$~\cite{Wang:2024fwc}&~&~ \\
\end{tabular}
\end{ruledtabular}
\end{table}

We firstly discuss how the numerical results of form factors are obtained. From the Eq. (\ref{eq:26}), the form factors depend on some input parameters such as the Borel parameters $T_1^2$ and $T_2^2$, the continuum threshold parameters $s_{0}$ and $u_{0}$, and the square momentum $Q^{2}$. To analyze the corresponding decay processes of $B_{c}$ to charmonia, the values of the form factors in time-like regions ($Q^{2}=-q^{2}<0$) need to be obtained. However, the calculations of three-point QCD sum rules are carried out in space-like regions ($Q^{2}>0$) or $Q^{2}=0$, because of the geometric constraints in Delta function integrals. Therefore, we uniformly get the values of form factors in space-like regions in this work, and extrapolating them into time-like regions.  

To obtain reliable sum rule results, two conditions should be satisfied, which are the pole dominance and convergence of OPE. The pole contribution is defined as~\cite{Shi:2019hbf},
\begin{eqnarray}\label{eq:37}
\notag
\mathrm{PC} = \frac{\int_{s_1}^{s_0} ds\int_{u_1}^{u_0} du\rho^{QCD} (s,u,Q^2)e^{-\frac{s}{T_1^2}-\frac{u}{T_2^2}}}{\int_{s_1}^\infty  ds\int_{u_1}^\infty  du\rho^{QCD}(s,u,Q^2)e^{-\frac{s}{T_1^2}-\frac{u}{T_2^2}}}\\
\end{eqnarray}

\begin{figure}
	\centering
	\includegraphics[width=8.5cm]{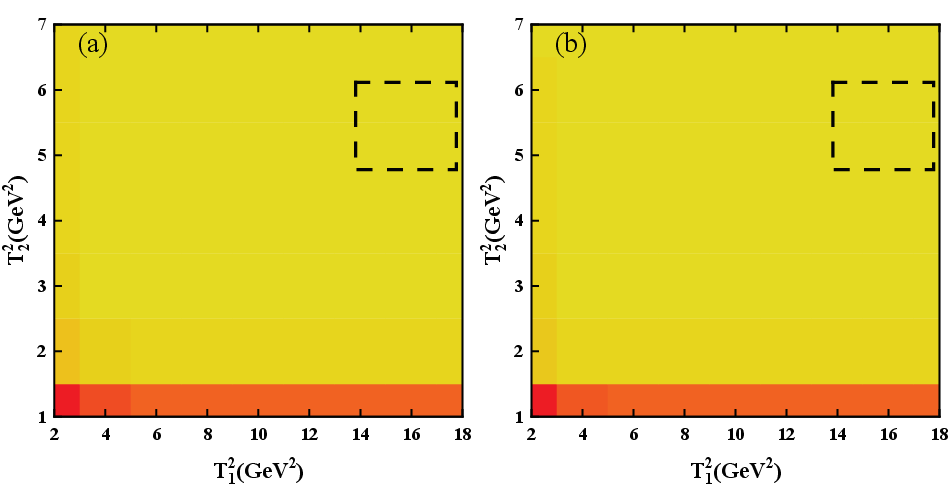}
	\caption{$F_{0}$ (a) and $F_{1}$ (b) in $Q^2=1$ GeV$^2$ as functions of the Borel parameters $T_1^2$ and $T_2^2$, where $T_1^2$ and $T_2^2$ are taken as free parameters. The larger the form factors, the darker the color. The 2D Borel platforms are enclosed by the dashed contours (the pole contribution in this region is about 75\%$\sim$55\%).}
	\label{FT1T2}
\end{figure}

The condition of pole dominance requires $\mathrm{PC}>50 \%$. An appropriate work region of Borel parameters should be selected to determine the values of form factors, where the final results are less dependent on the Borel parameters and at the same time the condition of pole dominance is also satisfied. This work region is commonly named as 'Borel platform'. Taking the axial vector form factors of $B_c \to \chi_{c0}$ as an example, by fixing $Q^2=1$ GeV$^2$ in Eq. (\ref{eq:26}), we plot the color figure of form factors $F_0$ and $F_1$ with variation of Borel parameters $T_1^2$ and $T_2^2$ which are shown in Fig. \ref{FT1T2}. The ranges of $T_1^2$ and $T_2^2$ are taken as $2\sim 18$ GeV$^2$ and $1\sim 7$ GeV$^2$, respectively. From these figures, one can find that the form factors will decrease as the Borel parameters increasing and tend to stabilize. Meantime, the pole contribution will also decrease as the Borel parameters increasing. By considering the conditions of pole dominance, the Borel platforms are determined and enclosed by the dashed contours in Fig. \ref{FT1T2}. The pole contribution in this region is about 75\%$\sim$55\%. It also can be found that the Borel parameters $T_1^2$ and $T_2^2$ approximately satisfy the relation $T_2^2/T_1^2\approx m_{\chi_{c0}}^2/m_{B_c}^2 \approx 0.3$ in this region. Thus, the relation $T_2^2=kT_1^2=kT^2$ mentioned above is feasible to reduce the double Borel parameters in our analysis.

\begin{figure}
	\centering
	\includegraphics[width=8.5cm]{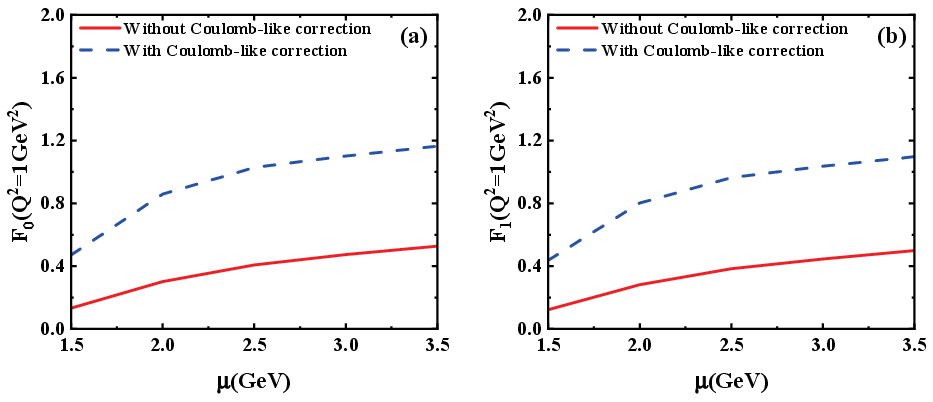}
	\caption{The form factors $F_0$ (a) and $F_{1}$ (b) in $Q^2=1$ GeV$^2$ with variation of energy scale $\mu$.}
	\label{Fmu}
\end{figure}

According to RGE in Eq. (\ref{eq:36}), the masses of the heavy quarks will change with the energy scale, which will affect the results of the dispersion integral in sum rules, which in turn will change the position of the Borel platform and the values of the form factors. Taking axial vector form factors $F_0$ and $F_1$ of $B_c \to \chi_{c0}$ as an example, we get the values of the form factors in $Q^2=1$ GeV$^2$ in different energy scale, which are shown in Fig. \ref{Fmu}. From these figures, one can find that the form factors increase with the increase of the energy scale, where the range of scale energy is $1.5\sim3.5$ GeV. In principle, if we only consider the leading order contribution, a suitable energy scale can not be obtained. We need to consider the higher order corrections to reduce the energy scale dependence of the results. Although we consider the Coulomb-like $\alpha_s/v$ correction in our analysis. However, the energy scale dependence of the results seem to not be significantly weakened. Therefore, we can only determine the energy scale according to phenomenological considerations. In Ref.~\cite{Wang:2024fwc}, we analyzed the mass and decay constant of $B_c$ meson in two-point QCD sum rules by considering the next-to-leading order contribution, and we got a suitable energy scale $\mu=2$ GeV for $B_c$ meson. In this study, we continue to use this energy scale to obtain the form factors and compare them with other theoretical models.  

 Taking $Q^{2}=1$ GeV$^{2}$, the Borel parameters are determined and Borel platform is chosen after repeated trial and contrast. The Borel platforms for all form factors are shown in Figs. \ref{BW1} and \ref{BW2} in Appendix \ref{Sec:AppA}. From Figs. \ref{BW1} and \ref{BW2}, we can see that the non-perturative contribution is tiny, the convergence of OPE is also satisfied.  The Borel platform, pole contributions in the Borel platform, and the values of form factors in $Q^{2}=1$ GeV$^{2}$ are listed in Table \ref{BPV}.
\begin{table}[htbp]
\renewcommand\arraystretch{1.5}
\begin{ruledtabular}\caption{The Borel platform (BP), pole contributions (PC) and the values of form factors (FF) in $Q^{2}=1$ GeV$^{2}$ for different decay modes. The star denote the results with Coulomb-like correction.}
\label{BPV}
\begin{tabular}{c c c c c c }
Mode&FF&BP (GeV$^{2}$)&PC(\%)&$F$&$F^{*}$ \\ \hline
\multirow{3}*{$B_{c}\to \chi_{c0}$}&$F_{0}$ &$15\sim17$&$64\sim58$&$0.30^{+0.02}_{-0.04}$&$0.84^{+0.06}_{-0.08}$ \\
~&$F_{1}$ &$15\sim17$&$63\sim57$&$0.28^{+0.02}_{-0.03}$&$0.78^{+0.06}_{-0.07}$ \\ 
~&$T$ &$18\sim20$&$73\sim69$&$0.35^{+0.03}_{-0.04}$&$0.99^{+0.07}_{-0.08}$  \\  \hline
\multirow{7}*{$B_{c}\to \chi_{c1}$}&$A$ &$19\sim21$&$71\sim67$&$0.19^{+0.02}_{-0.02}$&$0.53^{+0.04}_{-0.04}$ \\
~&$V_{0}$ &$23\sim25$&$79\sim76$&$0.060^{+0.013}_{-0.012}$&$0.15^{+0.03}_{-0.03}$  \\  
~&$V_{1}$ &$18\sim20$&$64\sim59$&$0.51^{+0.05}_{-0.06}$&$1.42^{+0.11}_{-0.13}$   \\
~&$V_{2}$ &$20\sim22$&$64\sim59$&$0.088^{+0.005}_{-0.007}$&$0.26^{+0.01}_{-0.02}$   \\
~&$T_{0}$ &$19\sim21$&$76\sim72$&$0.087^{+0.006}_{-0.008}$&$0.25^{+0.02}_{-0.02}$   \\
~&$T_{1}$ &$17\sim19$&$54\sim50$&$-0.44^{+0.05}_{-0.05}$&$-1.23^{+0.11}_{-0.13}$   \\
~&$T_{2}$ &$15\sim17$&$55\sim50$&$1.23^{+0.12}_{-0.15}$&$3.48^{+0.31}_{-0.34}$  \\ \hline
\multirow{7}*{$B_{c}\to h_{c}$}&$A$ &$9\sim11$&$55\sim46$&$0.075^{+0.010}_{-0.011}$&$0.19^{+0.03}_{-0.02}$ \\
~&$V_{0}$ &$13\sim15$&$67\sim61$&$0.80^{+0.07}_{-0.07}$&$2.30^{+0.16}_{-0.18}$   \\  
~&$V_{1}$ &$13\sim15$&$61\sim56$&$2.46^{+0.23}_{-0.26}$&$7.05^{+0.56}_{0.65}$   \\
~&$V_{2}$ &$9\sim11$&$57\sim50$&$0.16^{+0.02}_{-0.02}$&$0.46^{+0.05}_{-0.06}$   \\
~&$T_{0}$ &$16\sim18$&$90\sim87$&$0.16^{+0.01}_{-0.02}$&$0.45^{+0.03}_{-0.03}$  \\
~&$T_{1}$ &$13\sim15$&$77\sim72$&$-0.039^{+0.001}_{-0.002}$&$-0.13^{+0.01}_{-0.01}$   \\
~&$T_{2}$ &$13\sim15$&$84\sim78$&$0.49^{+0.02}_{-0.02}$&$1.48^{+0.04}_{-0.05}$  \\ \hline
\multirow{7}*{$B_{c}\to \chi_{c2}$}&$h$ &$18\sim20$&$70\sim65$&$0.031^{+0.003}_{-0.003}$&$0.089^{+0.005}_{-0.007}$  \\
~&$k$ &$13\sim15$&$76\sim69$&$1.91^{+0.14}_{-0.17}$&$5.40^{+0.34}_{-0.30}$  \\ 
~&$b_{+}$ &$13\sim15$&$64\sim55$&$-0.0099^{+0.0001}_{-0.0005}$&$-0.031^{+0.001}_{-0.001}$  \\
~&$b_{-}$ &$13\sim15$&$66\sim57$&$0.011^{+0.001}_{-0.001}$&$0.034^{+0.002}_{-0.001}$   \\ 
~&$T_{0}$ &$15\sim17$&$66\sim53$&$-0.085^{+0.007}_{-0.010}$&$-0.22^{+0.01}_{-0.03}$   \\ 
~&$T_{1}$ &$13\sim15$&$66\sim58$&$0.026^{+0.002}_{-0.002}$&$0.085^{+0.006}_{-0.007}$  \\
~&$T_{2}$ &$17\sim19$&$81\sim76$&$2.00^{+0.17}_{-0.18}$&$5.63^{+0.37}_{-0.43}$  \\
\end{tabular}
\end{ruledtabular}
\end{table}

It is noted that after considering the Coulomb-like correction, the values of the form factors is about three times of that without Coulomb-like correction. From Eq. (\ref{eq:28}), we can find that the Coulomb-like term relies on strong coupling constant $\alpha_s^C$. Doing the power series expansion on $\alpha_s^C$ of the Coulomb-like term in Eq. (\ref{eq:28}), we can obtain the following relation,
\begin{eqnarray}\label{eq:38}
	\notag
	&&\sqrt{\frac{4\pi \alpha _s^C}{3v_1}{\left[ 1 - \exp (-\frac{4\pi \alpha _s^C}{3v_1}) \right]}^{-1}} \sqrt{\frac{4\pi \alpha _s^C}{3v_2}{\left[ 1-\exp(- \frac{4\pi \alpha _s^C}{3v_2}) \right]}^{-1}} \\
	\notag
	&&= 1 + \alpha _s^C\left( \frac{\pi}{3v_1} + \frac{\pi }{3v_2} \right) + (\alpha _s^C)^2\left[\frac{\pi ^2}{54}{\left(\frac{1}{v_1} \right)^2} \right. + \frac{\pi ^2}{9}\frac{1}{v_1v_2}\\
	&&+ \left. \frac{\pi ^2}{54}\left( \frac{1}{v_2} \right)^2 \right] + O[(\alpha _s^C)^3] + ...
\end{eqnarray}
According to this equation, we can obtain the ratios among the perturbative contribution, leading order Coulomb-like correction ($\alpha_s^C$), next-to-leading order Coulomb-like correction (($\alpha_s^C$)$^2$) and the higher order contributions ($(\alpha_s^C)^n(n>2)$) for all form factors in $Q^2=1$ GeV$^2$, where the total Coulomb-like contributions are taken as 1 (see Table \ref{alphas}). From Table. \ref{alphas}, the contribution of the leading order Coulomb-like corrections for all form factors are about $1\sim1.5$ times as large as that of perturbative term, and the Coulomb-like corrections decrease quickly with increase of orders of $\alpha_s^C$. The finally results are convergent for the power expansion of $\alpha_s^C$. 
\begin{table}[htbp]
	\renewcommand\arraystretch{1.5}
	\begin{ruledtabular}\caption{The ratios among the perturbative contributions, leading order Coulomb-like corrections ($\alpha_s^C$), next-to-leading order Coulomb-like corrections (($\alpha_s^C$)$^2$) and the higher order contributions ($(\alpha_s^C)^n(n>2)$) for all form factors in $Q^2=1$ GeV$^2$, where the total Coulomb-like contributions are taken as 1.}
		\label{alphas}
		\begin{tabular}{c c c c c c }
			Mode&Form factor&Pert&$\alpha_s^C$&$(\alpha_s^C)^2$&$(\alpha_s^C)^n(n>2)$ \\ \hline
			\multirow{3}*{$B_{c}\to \chi_{c0}$}&$F_{0}$ &$0.35$&$0.49$&$0.11$&$0.05$ \\
			~&$F_{1}$ &$0.35$&$0.49$&$0.11$&$0.05$ \\ 
			~&$T$ &$0.35$&$0.49$&$0.11$&$0.05$  \\  \hline
			\multirow{7}*{$B_{c}\to \chi_{c1}$}&$A$ &$0.35$&$0.49$&$0.11$&$0.05$ \\
			~&$V_{0}$ &$0.40$&$0.45$&$0.066$&$0.064$  \\  
			~&$V_{1}$ &$0.36$&$0.48$&$0.10$&$0.06$   \\
			~&$V_{2}$ &$0.34$&$0.49$&$0.12$&$0.05$   \\
			~&$T_{0}$ &$0.35$&$0.49$&$0.11$&$0.05$   \\
			~&$T_{1}$ &$0.36$&$0.48$&$0.10$&$0.06$   \\
			~&$T_{2}$ &$0.35$&$0.48$&$0.10$&$0.07$  \\ \hline
			\multirow{7}*{$B_{c}\to h_{c}$}&$A$ &$0.39$&$0.46$&$0.076$&$0.074$ \\
			~&$V_{0}$ &$0.35$&$0.49$&$0.11$&$0.05$   \\  
			~&$V_{1}$ &$0.35$&$0.48$&$0.11$&$0.06$   \\
			~&$V_{2}$ &$0.34$&$0.47$&$0.093$&$0.097$   \\
			~&$T_{0}$ &$0.35$&$0.49$&$0.11$&$0.05$  \\
			~&$T_{1}$ &$0.30$&$0.51$&$0.15$&$0.04$   \\
			~&$T_{2}$ &$0.33$&$0.51$&$0.13$&$0.03$  \\ \hline
			\multirow{7}*{$B_{c}\to \chi_{c2}$}&$h$ &$0.35$&$0.49$&$0.11$&$0.05$  \\
			~&$k$ &$0.35$&$0.49$&$0.11$&$0.05$  \\ 
			~&$b_{+}$ &$0.32$&$0.53$&$0.15$&$0$  \\
			~&$b_{-}$ &$0.32$&$0.53$&$0.15$&$0$   \\ 
			~&$T_{0}$ &$0.39$&$0.45$&$0.09$&$0.07$   \\ 
			~&$T_{1}$ &$0.31$&$0.45$&$0.045$&$0.005$  \\
			~&$T_{2}$ &$0.36$&$0.49$&$0.11$&$0.04$  \\
		\end{tabular}
	\end{ruledtabular}
\end{table}

\begin{figure}
\centering
\includegraphics[width=8.5cm]{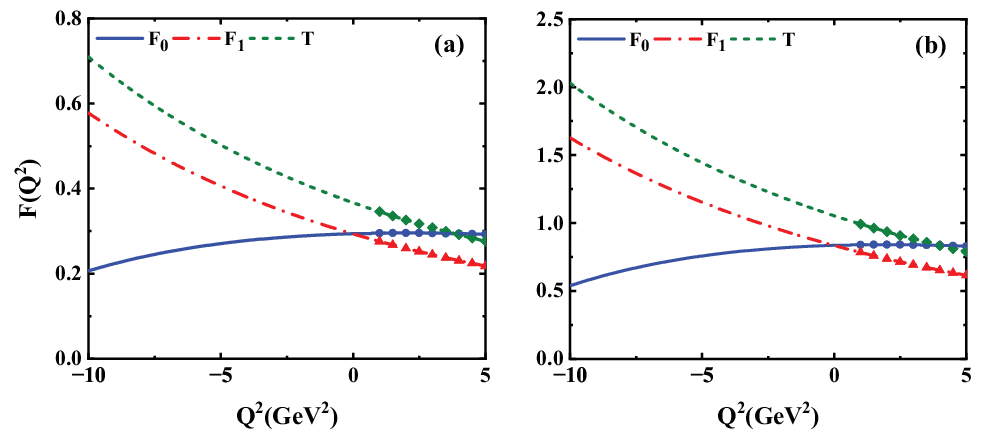}
\caption{Fitting results of form factors for $B_{c}\to \chi_{c0}$ without cosidering Coulomb-like correction (a) and considering Coulomb-like correction (b).}
\label{FFchic0}
\end{figure}

\begin{figure}
\centering
\includegraphics[width=8.5cm]{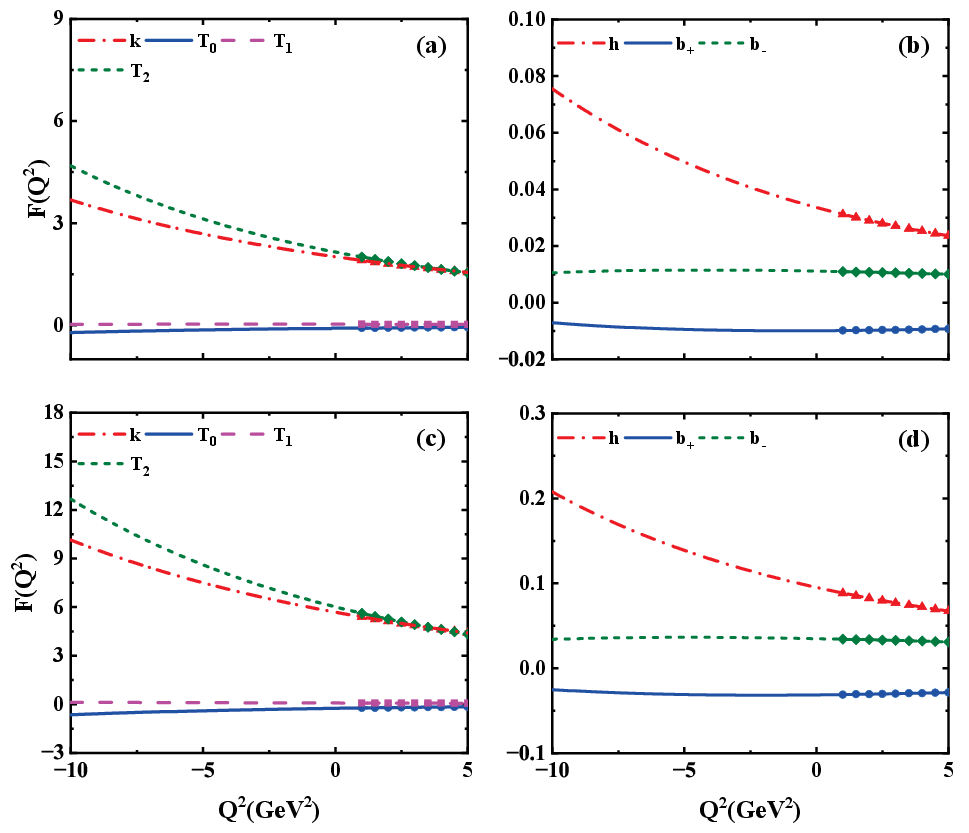}
\caption{Fitting results of form factors for $B_{c}\to \chi_{c2}$ without cosidering Coulomb-like correction (a, b) and considering Coulomb-like correction (c, d).}
\label{FFchic2}
\end{figure}

Taking different values of $Q^{2}$, the form factors in space-like regions ($Q^{2}>0$) can be obtained, where the range of $Q^{2}$ is taken as $1\sim5$ GeV$^{2}$, uniformly. The values of form factors in time-like regions are obtained by fitting the results in space-like regions with appropriate functions and by extrapolating these results into the time-like regions. The $z-$ series parameterization approach was proposed in Ref.~\cite{Boyd:1994tt}, and was widely employed to fit different form factors~\cite{Leljak:2019eyw,Bourrely:2008za,Zhao:2020mod,Wang:2015vgv,Cui:2022zwm}. With this method, the form factors can be expanded as the following series,
\begin{eqnarray}\label{eq:39}
\notag
F(Q^2) &&= \frac{1}{1 + Q^2/m_R^2} \\
&&\times \sum\limits_{k = 0}^{N - 1} {b_k\left[ z(Q^2,t_0)^k - {(- 1)^{k - N}}\frac{k}{N} \right.} \left. z(Q^2,t_0)^N \right]
\end{eqnarray}
where $m_{R}$ is the mass of low-lying $B_{c}$ resonance~\cite{Leljak:2019eyw} which is taken as 6.75 GeV in the present work. $b_{k}$ is the fitting parameter and the function $z(Q^{2},t_{0})$ is taken as the following form,    
\begin{eqnarray}\label{eq:40}
z(Q^2,t_0) &&= \frac{\sqrt {t_ + + Q^2}  - \sqrt {t_ +  - t_0} }{\sqrt {t_ +  + Q^2}  + \sqrt {t_ +  - t_0}} 
\end{eqnarray}
where $t_{\pm}=(m_{Bc}\pm m_{X})^{2}$, and $t_{0}=t_{+}-\sqrt{t_{+}(t_{+}-t_{-})}$~\cite{Leljak:2019eyw,Cui:2022zwm}. The $z-$ series are truncated at $N=3$ in our calculation, and the values of fitting parameters $b_{k}$ for all form factors are listed in Table \ref{ZP}. The fitting diagrams of these form factors are shown in Figs. \ref{FFchic0}$\sim$\ref{FFchic2}.

\begin{table}[htbp]
\begin{ruledtabular}\caption{Fitting parameters of different form factors (FF) in $z-$ series expand approach. $b_{k}^{*}$ denote the fitting parameters with Coulomb-like correction.}
\label{ZP}
\begin{tabular}{c c c c c c c c c}
Mode&FF&$b_{0}$&$b_{1}$&$b_{2}$&$b_{0}^{*}$&$b_{1}^{*}$&$b_{2}^{*}$ \\ \hline
\multirow{3}*{$B_{c}\to \chi_{c0}$}&$F_{0}$ &$0.25$&$4.3$&$-53$&$0.70$&$14$&$-200$ \\
~&$F_{1}$ &$0.35$&$-5.2$&$34$&$0.99$&$-14$&$75$ \\ 
~&$T$ &$0.43$&$-6.2$&$43$&$1.24$&$-17$&$97$  \\ \hline
\multirow{7}*{$B_{c}\to \chi_{c1}$}&$A$ &$0.23$&$-3.3$&$24$&$0.65$&$-8.8$&$48$ \\
~&$V_{0}$ &$0.073$&$-0.94$&$4.2$&$0.19$&$-2.7$&$14$ \\ 
~&$V_{1}$ &$0.42$&$9.0$&$-96$&$1.2$&$26$&$-310$  \\
~&$V_{2}$ &$0.098$&$-0.71$&$2.9$&$0.29$&$-2.1$&$2.4$ \\ 
~&$T_{0}$ &$0.11$&$-2.0$&$17$&$0.32$&$-5.7$&$47$  \\ 
~&$T_{1}$ &$-0.55$&$8.9$&$-65$&$-1.6$&$26$&$-190$ \\ 
~&$T_{2}$ &$1.5$&$-18$&$101$&$4.2$&$-55$&$320$  \\  \hline
\multirow{7}*{$B_{c}\to h_{c}$}&$A$ &$0.099$&$-1.9$&$14$&$0.26$&$-5.2$&$37$ \\
~&$V_{0}$ &$1.0$&$-16$&$110$&$2.9$&$-48$&$330$ \\ 
~&$V_{1}$ &$2.9$&$-36$&$200$&$8.6$&$-120$&$750$  \\
~&$V_{2}$ &$0.20$&$-3.5$&$22$&$0.61$&$-13$&$96$ \\ 
~&$T_{0}$ &$0.21$&$-4.2$&$38$&$0.60$&$-12$&$110$  \\ 
~&$T_{1}$ &$-0.044$&$0.26$&$7.7$&$-0.18$&$3.9$&$-37$ \\ 
~&$T_{2}$ &$0.69$&$-16$&$130$&$2.1$&$-51$&$420$  \\  \hline
\multirow{7}*{$B_{c}\to \chi_{c2}$}&$h$ &$0.041$&$-0.80$&$6.6$&$0.12$&$-2.4$&$24$ \\
~&$k$ &$2.3$&$-26$&$149$&$6.3$&$-66$&$274$ \\ 
~&$b_{+}$ &$-0.0088$&$-0.15$&$3.2$&$-0.028$&$-0.39$&$9.5$  \\
~&$b_{-}$ &$0.010$&$0.10$&$-2.6$&$0.033$&$0.29$&$-8.5$ \\ 
~&$T_{0}$ &$-0.12$&$2.4$&$-13$&$-0.31$&$8.0$&$-77$  \\ 
~&$T_{1}$ &$0.025$&$0.30$&$-11$&$0.093$&$-0.34$&$-11$ \\ 
~&$T_{2}$ &$2.6$&$-46$&$407$&$7.1$&$-118$&$890$  \\ 
\end{tabular}
\end{ruledtabular}
\end{table}

\begin{figure}
\centering
\includegraphics[width=8.5cm]{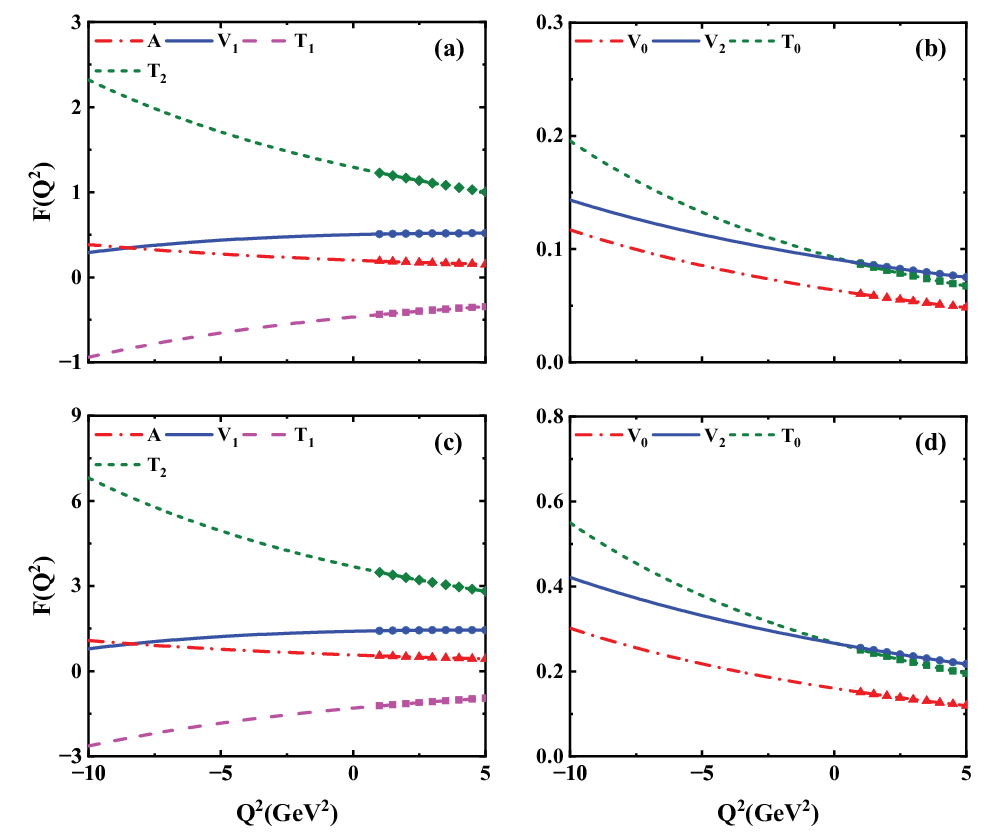}
\caption{Fitting results of form factors for $B_{c}\to \chi_{c1}$ without considering Coulomb-like correction (a, b) and considering Coulomb-like correction (c, d).}
	\label{FFchic1}
\end{figure} 

\begin{figure}
\centering
\includegraphics[width=8.5cm]{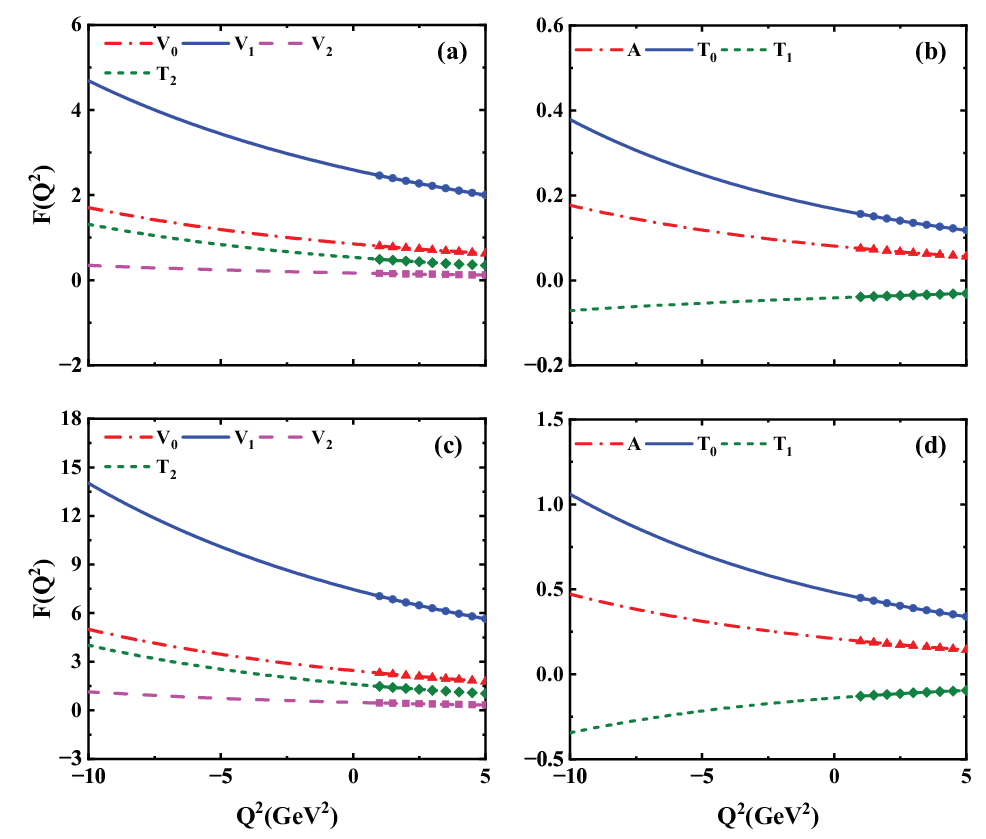}
\caption{Fitting results of form factors for $B_{c}\to h_{c}$ without considering Coulomb-like correction (a, b) and considering Coulomb-like correction (c, d).}
\label{FFhc}
\end{figure} 

The values of form factors at $Q^{2}=0$ are shown in Table \ref{FF}. As a contrast, the results of other collaborations are also shown in this table. One can find that the form factors with Coulomb-like correction are about three times that of no Coulomb-like correction. For the axial vector form factors $F_{0,1}$ of $B_{c}\to \chi_{c0}$, the axial vector form factor $A$ and vector form factor $V_{2}$ of $B_{c}\to \chi_{c1}$, our results with no Coulomb-like correction are consistent well with those calculated using the Light-front quark model (LFQM)~\cite{Zhang:2023ypl,Li:2023wgq,Hazra:2023zdk}, but more small for the results given by other LFQM~\cite{Wang:2009mi}. For the vector form factors $V_0$ and $V_1$ of $\chi_{c1}$, our results are well consistent with Ref.~\cite{Hazra:2023zdk}. After the Coulomb-like correction is considered, our results are larger than their predictions but smaller than the NRQCD predictions~\cite{Zhu:2017lwi}. For the transitions of $B_{c}\to \chi_{c2}$, our predictions of $h$, $b_{+,-}$ without Coumlomb-like correction are compatible with Ref.~\cite{Wang:2009mi}, but too larger than Ref.~\cite{Azizi:2013zta}. For other form factors, various theory predictions are not consistent with each other especially for $B_{c}\to h_{c}$ transitions. In Table \ref{FF}, we can find that our result for the vector form factor $V_{1}$ of $B_{c}\to h_{c}$ is too larger than other predictions. Moreover, the results for the vector form factor $V_{2}$ are negative in LFQM and NRQCD calculations, but they are positive in QCD sum rules predictions. From Table \ref{FF}, one can also find that our results are not well compatible with other sum rules predictions. By detailed analysis, we find the main discrepancies originate from different values of the decay constant for hadron, of the energy scale and the type of interpolating currents of $P$-wave charmoina etc. In Ref.~\cite{Azizi:2009ny}, the vector and axial vector form factors of $B_{c}\to \chi_{c0,1}$ and $h_{c}$ are analyzed with three-point QCD sum rules. In this article, the central values of decay constants for $B_{c}$ and all $P$-wave charmoina are taken as $0.4$ GeV and $0.34$ GeV, the energy scale of heavy quarks is taken as $m_{c}$. In addition, the axial vector current $\bar{c}\gamma_{\mu}\gamma_{5}c$ is selected to interpolate both $\chi_{c1}$ and $h_{c}$. In our present work, the decay constants of $B_{c}$ is taken as $0.371\pm0.037$ GeV which originates from two-point QCD sum rules calculation considering the next-to-leading contribution of perturbative term~\cite{Wang:2024fwc}, the values of decay constants for $\chi_{c0}$, $\chi_{c1}$ and $h_{c}$ are taken from Refs.~\cite{Novikov:1977dq,Becirevic:2013bsa} which are $0.343$, $0.338$ and $0.235$ GeV, the heavy quark energy scale $\mu=2$ GeV is also taken from Ref.~\cite{Wang:2024fwc}. In the present work, we choose the tensor current $\bar{c}\sigma_{\mu\nu}c$ instead of axial vector current to interpolate $h_{c}$ because the quantum number $J^{PC}$ of $h_{c}$ is $1^{+-}$, but the $C$ parity of axial current is $+$~\cite{Wang:2012gj}. In Ref.~\cite{Azizi:2013zta}, the vector and axial vector form factors of $B_{c}\to \chi_{c2}$ are analyzed by three-point QCD sum rules, the authors get the QCD spectral density by using Feynman parameterization approach in QCD side. In our analysis, the QCD spectral density is obtained by Cutkoskys's rules. This is the main reason for the disagreement. 

\begin{table*}[htbp]
\renewcommand\arraystretch{1.5}
\begin{ruledtabular}\caption{The values of the form factors at $Q^{2}=0$. 'This work$^{*}$' denotes the results with Coulomb-like correction.}
\label{FF}
\begin{tabular}{c c c c c c c c c c}
Modes&Form factors&This work&This work$^{*}$&~\cite{Azizi:2009ny,Azizi:2013zta} &~\cite{Wang:2009mi}&~\cite{Zhang:2023ypl}&~\cite{Li:2023wgq}&~\cite{Hazra:2023zdk}&~\cite{Zhu:2017lwi} \\ \hline
\multirow{3}*{$B_{c}\to \chi_{c0}$}&$F_{0}$ &$0.29^{+0.03}_{-0.03}$&$0.84^{+0.06}_{-0.08}$&$0.67$&$0.47$&$0.33$&$0.337$&$-$&$1.65$ \\
~&$F_{1}$ &$0.29^{+0.03}_{-0.03}$&$0.83^{+0.07}_{-0.07}$&$0.67$&$0.47$&$0.33$&$0.337$&$-$&$1.65$\\ 
~&$T$ &$0.37^{+0.03}_{-0.04}$&$1.05^{+0.07}_{-0.08}$&$-$&$-$&$-$&$-$&$-$&$-$ \\ \hline
\multirow{7}*{$B_{c}\to \chi_{c1}$}&$A$ &$0.20^{+0.02}_{-0.02}$&$0.57^{+0.03}_{-0.05}$&$0.13$&$0.36$&$0.24$&$0.215$&$0.21$&$1.24$ \\
~&$V_{0}$ &$0.064^{+0.013}_{-0.012}$&$0.16^{+0.03}_{-0.03}$&$0.03$&$0.13$&$0.15$&$0.025$&$0.07$&$0.17$\\ 
~&$V_{1}$ &$0.50^{+0.05}_{-0.55}$&$1.40^{+0.12}_{-0.12}$&$0.28$&$0.85$&$0.72$&$0.339$&$0.47$&$2.87$ \\ 
~&$V_{2}$ &$0.091^{+0.005}_{-0.006}$&$0.27^{+0.01}_{-0.02}$&$0.059$&$0.15$&$0.10$&$0.078$&$0.08$&$0.57$\\ 
~&$T_{0}$ &$0.093^{+0.006}_{-0.008}$&$0.27^{+0.01}_{-0.02}$&$-$&$-$&$-$&$-$&$-$&$-$ \\
~&$T_{1}$ &$-0.47^{+0.06}_{-0.06}$&$-1.31^{+0.11}_{-0.14}$&$-$&$-$&$-$&$-$&$-$&$-$\\ 
~&$T_{2}$ &$1.29^{+0.13}_{-0.14}$&$3.68^{+0.31}_{-0.35}$&$-$&$-$&$-$&$-$&$-$&$-$ \\ \hline
\multirow{7}*{$B_{c}\to h_{c}$}&$A$ &$0.081^{+0.010}_{-0.012}$&$0.21^{+0.02}_{-0.02}$&$0.14$&$0.07$&$0.06$&$0.039$&$0.04$&$0.07$ \\
~&$V_{0}$ &$0.85^{+0.08}_{-0.07}$&$2.46^{+0.16}_{-0.18}$&$0.03$&$0.64$&$0.41$&$0.390$&$0.35$&$2.11$\\ 
~&$V_{1}$ &$2.59^{+0.24}_{-0.26}$&$7.47^{+0.57}_{-0.65}$&$0.29$&$0.50$&$0.42$&$0.298$&$0.30$&$0.54$ \\ 
~&$V_{2}$ &$0.17^{+0.02}_{-0.02}$&$0.49^{+0.07}_{-0.05}$&$0.059$&$-0.32$&$-0.18$&$-0.196$&$-0.16$&$-0.99$\\ 
~&$T_{0}$ &$0.17^{+0.01}_{-0.02}$&$0.48^{+0.03}_{-0.03}$&$-$&$-$&$-$&$-$&$-$&$-$ \\
~&$T_{1}$ &$-0.041^{+0.002}_{-0.002}$&$-0.14^{+0.00}_{-0.01}$&$-$&$-$&$-$&$-$&$-$&$-$\\ 
~&$T_{2}$ &$0.54^{+0.01}_{-0.03}$&$1.62^{+0.04}_{-0.05}$&$-$&$-$&$-$&$-$&$-$&$-$ \\ \hline
\multirow{7}*{$B_{c}\to \chi_{c2}$}&$h$ &$0.034^{+0.002}_{-0.003}$&$0.095^{+0.005}_{-0.007}$&$-1.70\times10^{-4}$&$0.022$&$-$&$-$&$-$&$-$ \\
~&$k$ &$2.06^{+0.10}_{-0.21}$&$5.67^{+0.37}_{-0.38}$&$0.18$&$1.27$&$-$&$-$&$-$&$-$\\ 
~&$b_{+}$ &$-0.010^{+0.001}_{-0.004}$&$-0.031^{+0.002}_{-0.001}$&$-0.038$&$-0.011$&$-$&$-$&$-$&$-$ \\ 
~&$b_{-}$ &$0.011^{+0.001}_{-0.000}$&$0.035^{+0.001}_{-0.001}$&$-0.056$&$0.020$&$-$&$-$&$-$&$-$\\ 
~&$T_{0}$ &$-0.097^{+0.003}_{-0.015}$&$-0.24^{+0.01}_{-0.03}$&$-$&$-$&$-$&$-$&$-$&$-$ \\
~&$T_{1}$ &$0.027^{+0.002}_{-0.003}$&$0.089^{+0.006}_{-0.007}$&$-$&$-$&$-$&$-$&$-$&$-$\\ 
~&$T_{2}$ &$2.18^{+0.14}_{-0.22}$&$6.02^{+0.38}_{-0.42}$&$-$&$-$&$-$&$-$&$-$&$-$ \\
\end{tabular}
\end{ruledtabular}
\end{table*}

\subsection{Numerical results of decay widths and branching ratios}
With these above form factors, the decay widths and branching ratios of the semileptonic and nonleptonic decays related to $B_{c}$ to $P$-wave charmonia transition are calculated in this section. The relevant input parameters are listed in Table \ref{IP2}. 

\begin{figure}
	\centering
	\includegraphics[width=8.5cm]{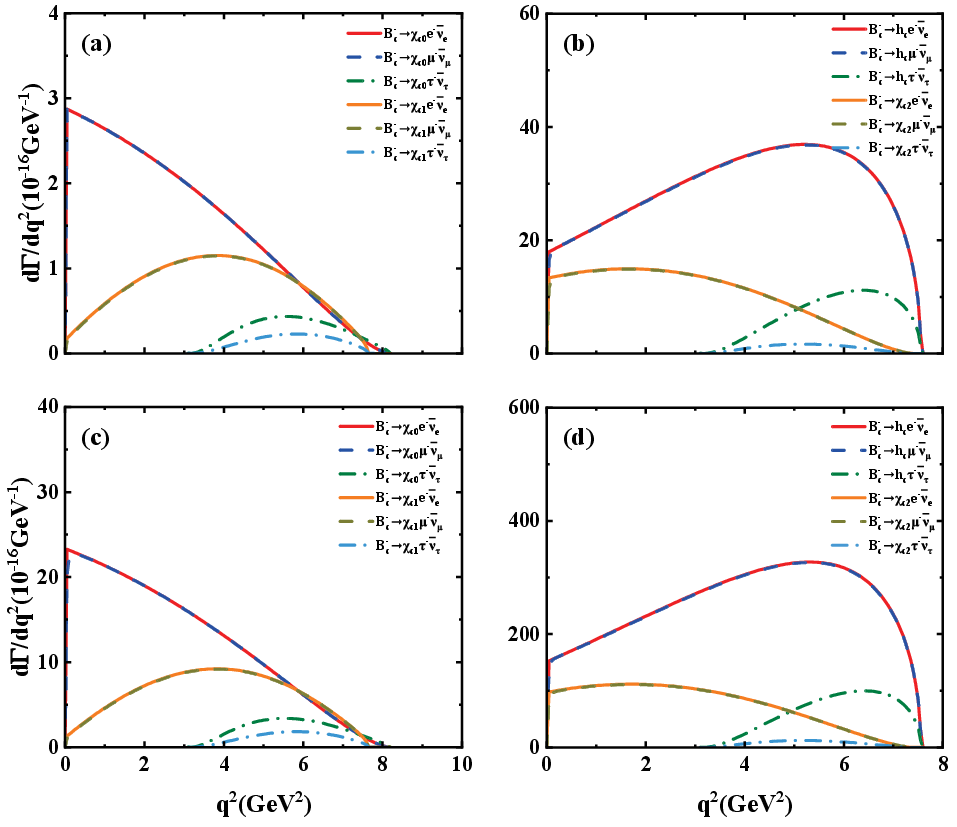}
	\caption{The differential decay width $d\Gamma/dq^{2}$ with variations of $q^{2}$ for semileptonic decay processes $B_{c}\to \chi_{c0}l\nu_{l}$, $\chi_{c1}l\nu_{l}$ and $B_{c}\to h_{c}l\nu_{l}$, $\chi_{c2}l\nu_{l}$ without considering the Coulomb-like correction ((a) and (b)) and considering the Coulomb-like correction ((c) and (d)).}
	\label{DW}
\end{figure}

\begin{table}[htbp]
\begin{ruledtabular}\caption{Input parameters (IP) in analysis the decay processes of $B_{c}$ meson.}
\label{IP2}
\begin{tabular}{c c c c }
IP&Values~\cite{ParticleDataGroup:2024cfk}&IP&Values \\ \hline
$m_{\pi}$&0.140 GeV&$f_{K}$&0.160 GeV~\cite{ParticleDataGroup:2024cfk}\\
$m_{K}$&0.494 GeV&$f_{\rho}$&0.216 GeV~\cite{ParticleDataGroup:2024cfk} \\
$m_{\rho}$&0.775 GeV&$f_{K^{*}}$&0.217 GeV~\cite{ParticleDataGroup:2024cfk} \\
$m_{K^{*}}$&0.892 GeV&$V_{cb}$&0.041~\cite{ParticleDataGroup:2024cfk} \\
$m_{e}$&0.511 MeV&$V_{us}$&0.224~\cite{ParticleDataGroup:2024cfk} \\
$m_{\mu}$&105.7 MeV&$V_{ud}$&0.974~\cite{ParticleDataGroup:2024cfk} \\
$m_{\tau}$&1.78 GeV&$a_{1}$&1.07~\cite{Buchalla:1995vs} \\
$f_{\pi}$&0.131 GeV&~&~ \\
\end{tabular}
\end{ruledtabular}
\end{table}
For three-body semileptonic decay $B_{c}\to Xl\bar{\nu}_{l}$ $(l=e, \mu$ and $\tau)$, the differential decay width can be expressed as,
\begin{eqnarray}\label{eq:41}
	\notag
	d\Gamma (B_c \to Xl\bar{\nu} _l) &&= \frac{1}{2J + 1}\sum \frac{1}{2m_{B_c}}d\Phi (p \to p',l,v)|T|^2 \\
	\notag
	d\Phi (p \to p',l,v) &&= (2\pi )^4\delta ^4(p - p' - l - v)\\
	&&\times \frac{d^3\vec p'}{(2\pi )^32p'_0}\frac{d^3\vec l}{(2\pi )^32l_0}\frac{d^3\vec v}{(2\pi )^32v_0}
\end{eqnarray}
where $J$ is the total angular momentum of $B_{c}$ meson, $\Sigma$ denotes the summation of all the polarization, $d\Phi(p\to p', l, v)$ is the three-body phase space, $p$, $p'$, $l$ and $v$ are four momentum for $B_{c}$, $X$, $l$ and $\bar{\nu}_{l}$ and $T$ denotes the transition matrix. By inserting intermediate state momentum $q=l+v$, the three-body phase space can be decomposed as product of two-body phase space,
\begin{eqnarray}\label{eq:42}
	d\Phi (p \to p',l,v) = \frac{dq^2}{2\pi }d\Phi (p \to p',q)d\Phi (q \to l,v)
\end{eqnarray}
where the two-body space phase $d\Phi(p\to p',q)$ can be written as,
\begin{eqnarray}\label{eq:43}
	\notag
	d\Phi (p \to p',q) &&= (2\pi )^4\delta ^4(p - p' - q) \frac{d^3\vec p'}{(2\pi )^3 2p'_{0}}\frac{d^3\vec q}{(2\pi )^3 2q_{0}}\\
\end{eqnarray} 
With the Eqs. (\ref{eq:31}), (\ref{eq:32}) and (\ref{eq:41}) $\sim$ (\ref{eq:43}), the differential decay widths of $B_{c}\to Xl\bar{\nu}_{l}$ can be obtained. The full expressions are shown in Appendix \ref{Sec:AppB}. The differential decay widths with variations of $q^2$ are shown in Fig. \ref{DW}. The decay widths and branching ratios for the three-body semileptonic decays can be derived by finishing the integration of $q^2$. The numerical results are collected in Table \ref{DWBR}. From this table, one can find that $\Gamma(B_{c}\to h_{c}l\bar{\nu}_{l})>\Gamma(B_{c}\to \chi_{c2}l\bar{\nu}_{l})>\Gamma(B_{c}\to \chi_{c0}l\bar{\nu}_{l})>\Gamma(B_{c}\to \chi_{c1}l\bar{\nu}_{l})$ in our prediction, the decay processes of $B_{c}\to h_{c}l\bar{\nu}_{l}$ and $B_{c}\to \chi_{c2}l\bar{\nu}_{l}$ have large branching ratios. Comparing with results in Refs.~\cite{Azizi:2009ny,Azizi:2013zta,Ivanov:2006ni,Hernandez:2006gt,Wang:2009mi,Li:2023wgq,Chang:2001pm} for the $B_{c}\to \chi_{cJ}l\bar{\nu}_{l}(J=0,1,2)$ and $h_{c}l\bar{\nu}_{l}$ decays which
are also collected in Table \ref{DWBR}, we can see that most of our results for branching ratios of $B_{c}\to \chi_{c0}l\bar{\nu}_{l}$ and $B_{c}\to \chi_{c1}l\bar{\nu}_{l}$ without Coulomb-like correction are comparable with their predictions. For the branching ratios of $B_{c}\to h_{c}l\bar{\nu}_{l}$ and $B_{c}\to \chi_{c2}l\bar{\nu}_{l}$, our predictions are larger than other collaborations. When we consider the Coulomb-like correction, these results are much larger than other predictions. This is due to the fact that the form factors will be about three times larger after considering the Coulomb-like correction, which will result in about nine times the decay widths and branching ratios. As the next leading order contribution, the Coulomb-like correction is too large. Thus, the results without Coulomb-like correction seem more reasonable. In conclusion, the rigorous calculation of next-to-leading order correction for perturbative contribution of form factors eagerly need to be carried out, and the discrepancies of different theoretical results will be tested by more experiments in the future.

\begin{table*}[htbp]
	\renewcommand\arraystretch{1.5}
	\begin{ruledtabular}
		\caption{Decay widths (in $10^{-6}$eV) and branching ratios ($\%$) of $B_{c}$ semileptonic decays. Branching ratios are calculated at $\tau_{B_c}$ = 0.51 ps~\cite{ParticleDataGroup:2024cfk}. The superscript star denotes the results obtained by considering Coulomb-like correction.}
		\label{DWBR}
		\begin{tabular}{c >{\centering}p{7em}| c c | c c c c c c c c c}
			&\multirow{2}{*}{Decay channels} & \multicolumn{2}{c|}{Decay widths} & \multicolumn{7}{c}{Branching ratios}\\
			&&This work & This work$^*$ &This work & This work$^*$&~\cite{Wang:2009mi}&~\cite{Chang:2001pm}&~\cite{Ivanov:2006ni}&~\cite{Hernandez:2006gt}&~\cite{Azizi:2009ny,Azizi:2013zta}&~\cite{Li:2023wgq}&~\cite{Hazra:2023zdk}\\
			\hline
			&$B_c^- \to \chi_{c0} e \bar{\nu}_{e}$ & $1.25^{+0.20}_{-0.24}$ & $9.99^{+1.29}_{-1.44}$ & $0.097^{+0.016}_{-0.019}$ &$0.78^{+0.09}_{-0.12}$&$0.21$&$0.12$&$0.17$&$0.11$&$0.182$&$0.11$&$-$\\
			&$B_c^- \to \chi_{c0} \mu \bar{\nu}_{\mu}$ & $1.24^{+0.20}_{-0.24}$ & $9.93^{+1.26}_{-1.45}$ & $0.096^{+0.016}_{-0.018}$ &  $0.77^{+0.10}_{-0.11}$&$0.21$&$0.12$&$0.17$&$0.11$&$0.182$&$0.11$&$-$\\
			&$B_c^- \to \chi_{c0} \tau \bar{\nu}_{\tau}$ & $0.14^{+0.02}_{-0.02}$ & $1.05^{+0.20}_{-0.08}$ & $0.011^{+0.002}_{-0.002}$ &  $0.082^{+0.015}_{-0.007}$&$0.024$&$0.017$&$0.013$&$0.013$&$0.049$&$0.0094$&$-$\\
			&$B_c^- \to \chi_{c1} e \bar{\nu}_{e}$ & $0.61^{+0.15}_{-0.10}$ & $4.85^{+0.71}_{-0.83}$ & $0.047^{+0.012}_{-0.007}$ &  $0.38^{+0.05}_{-0.07}$&$0.14$&$0.15$&$0.092$&$0.066$&$0.146$&$0.030$&$0.052$\\
			&$B_c^- \to \chi_{c1} \mu \bar{\nu}_{\mu}$ & $0.60^{+0.15}_{-0.09}$ & $4.82^{+0.70}_{-0.82}$ & $0.047^{+0.011}_{-0.007}$ &  $0.37^{+0.06}_{-0.06}$&$0.14$&$0.15$&$0.092$&$0.066$&$0.146$&$0.029$&$0.052$\\
			&$B_c^- \to \chi_{c1} \tau \bar{\nu}_{\tau}$ & $0.064^{+0.018}_{-0.007}$ & $0.51^{+0.08}_{-0.08}$ & $0.0049^{+0.0014}_{-0.0005}$ &  $0.039^{+0.007}_{-0.005}$&$0.015$&$0.024$&$0.0089$&$0.0072$&$0.0147$&$0.0026$&$0.0057$\\
			&$B_c^- \to h_{c} e \bar{\nu}_{e}$ & $22.15^{+3.18}_{-3.43}$ & $194.23^{+23.08}_{-24.12}$ & $1.72^{+0.24}_{-0.27}$ &  $15.06^{+1.79}_{-1.87}$&$0.31$&$0.18$&$0.27$&$0.17$&$0.142$&$0.11$&$0.105$\\
			&$B_c^- \to h_{c} \mu \bar{\nu}_{\mu}$ & $22.05^{+3.16}_{-3.41}$ & $193.31^{+22.98}_{-23.96}$ & $1.71^{+0.24}_{-0.27}$ &  $14.98^{+1.79}_{-1.85}$&$0.31$&$0.18$&$0.27$&$0.17$&$0.142$&$0.11$&$0.104$\\
			&$B_c^- \to h_{c} \tau \bar{\nu}_{\tau}$ & $3.03^{+0.36}_{-0.46}$ & $26.92^{+3.08}_{-2.66}$ & $0.24^{+0.02}_{-0.04}$ &  $2.09^{+0.24}_{-0.21}$&$0.022$&$0.025$&$0.017$&$0.015$&$0.0137$&$0.0051$&$0.0088$\\
			&$B_c^- \to \chi_{c2} e \bar{\nu}_{e}$ & $7.50^{+0.79}_{-0.81}$ & $55.92^{+7.83}_{-6.64}$ & $0.58^{+0.06}_{-0.06}$ &  $4.34^{+0.51}_{-0.52}$&$0.17$&$0.19$&$0.17$&$0.13$&$0.130$&$-$&$-$\\
			&$B_c^- \to \chi_{c2} \mu \bar{\nu}_{\mu}$ & $7.43^{+0.78}_{-0.80}$ & $55.38^{+7.74}_{-6.58}$ & $0.57^{+0.07}_{-0.06}$ &  $4.29^{+0.60}_{-0.51}$&$0.17$&$0.19$&$0.17$&$0.13$&$0.130$&$-$&$-$\\
			&$B_c^- \to \chi_{c2} \tau \bar{\nu}_{\tau}$ & $0.41^{+0.04}_{-0.01}$ & $3.01^{+0.44}_{-0.29}$ & $0.031^{+0.004}_{-0.001}$ &  $0.23^{+0.04}_{-0.02}$&$0.0092$&$0.029$&$0.0082$&$0.0093$&$0.020$&$-$&$-$\\
		\end{tabular}
	\end{ruledtabular}
\end{table*}

For two-body nonleptonic decay $B_{c}\to XP[V]$, the decay width can be expressed by the following standard two-body decay formula,
\begin{eqnarray}\label{eq:44}
\Gamma  &&= \frac{1}{2J + 1}\sum\int \frac{1}{2m_{B_c}}d\Phi (p \to p',q) |T|^2
\end{eqnarray}
With the Eqs. (\ref{eq:34}), (\ref{eq:35}) and (\ref{eq:44}), the decay widths of $B_{c}\to X P[V]$ can be obtained. The full expressions of decay widths are shown in Appendix \ref{Sec:AppC}. The numerical results of these two-body decay widths and branching ratios and those of other collaboration's are explicitly shown in Table \ref{TBD}. It can be seen from Table \ref{TBD} that the theoretical results of different collaborations are not consistent well with each other. The experimental data for the weak decay of $B_{c}$ to $P$-wave charmonia are scarce. Only the branching ratios of decay channel $B_{c}\to\chi_{c0}\pi$ is given as $\frac{\sigma(B_c^+)}{\sigma(B^+)}\times Br(B_{c}^{+}\to \chi_{c0}\pi^{+})=(9.8^{+3.4}_{-3.0}$$($stat$)\pm$0.8$($syst$))\times 10^{-6}$ by the LHCb collaboration~\cite{LHCb:2014mvo}. With our results $Br(B_{c}\to\chi_{c0}\pi)=0.24^{+0.05}_{-0.04}\times10^{-3}$ without Coulomb-like correction and $1.96^{+0.21}_{-0.30}\times10^{-3}$ with Coulomb-like correction, the cross section ration $\frac{\sigma(B_c^+)}{\sigma(B^+)}$ can be extracted to be about $(2\sim7)\times10^{-2}$ and $(3\sim8)\times10^{-3}$, respectively, which is useful to study the total cross section of $B_c$ meson. Recently, a systematic analysis about the $B_c$ nonleptonic decay ratios are obtained by using the factorization approach\cite{Losacco:2023uvp}. For pseudoscalar mesons $\pi$ and $K$, their values about the decay ratios are $B^{\chi_{c0}}_{\chi_{c2}}(\pi)=0.658$, $B^{h_c}_{\chi_{c2}}(\pi)=1.597$, $B^{\chi_{c0}}_{\chi_{c2}}(K)=0.663$ and $B^{h_c}_{\chi_{c2}}(K)=1.645$, where $B^{\chi_{c0}}_{\chi_{c2}}(\pi)$ denotes $\frac{Br(B_c\to \chi_{c0}\pi)}{Br(B_c\to \chi_{c2}\pi)}$. Our results are $0.159 (0.178)$, $1.238(1.405)$, $0.172(0.183)$ and $1.273(1.463)$, where the values in parentheses are obtained by considering the Coulomb-like correction. It can be found that our predictions for $B^{\chi_{c0}}_{\chi_{c2}}(\pi/K)$ are smaller, while the results for $B^{h_c}_{\chi_{c2}}(\pi/K)$ are compatible with theirs. For vector mesons $\rho$ and $K^*$, their values about the decay ratios are $B^{\chi_{c1}}_{\chi_{c0}}(\rho)=0.206$, $B^{\chi_{c0}}_{\chi_{c2}}(\rho)=0.590$, $B^{h_c}_{\chi_{c2}}(\rho)=1.312$, $B^{\chi_{c1}}_{\chi_{c0}}(K^*)=0.276$, $B^{\chi_{c0}}_{\chi_{c2}}(K^*)=0.570$ and $B^{h_c}_{\chi_{c2}}(K^*)=1.231$. Our results are $0.106(0.101)$, $0.156(0.172)$, $1.163(1.350)$, $0.124(0.119)$, $0.150(0.169)$ and $1.227(1.413)$, respectively. It is indicted that our results for $B^{\chi_{c1}}_{\chi_{c0}}(\rho/ K^*)$ and $B^{\chi_{c0}}_{\chi_{c2}}(\rho/ K^*)$ are smaller, while the results for $B^{h_c}_{\chi_{c2}}(\rho/K^*)$ are close to theirs. In short, both predictions show the same conclusion that the nonleptonic decay width of $B_c$ to $P$-wave charmonia satisfy the relation $\chi_{c2}>h_c>\chi_{c0}>\chi_{c1}$, while the predictions for ratios are different with each other. Overall, further experimental measurements are eagerly awaited to testify the discrepancies of different theoretical predictions.

It is noted that the NFA is adopted to analyze the two-body nonleptonic decays in our present work. However, as the simplest model to study the two-body weak decays of the heavy flavor hadron, NFA does not take into account the hard gloun exchange between the energetic meson emitted from the weak vertex and the $B_{c}$ transition form factors. Thus, the two-body decay width formulas in the framework of NFA are not renormalization invariant, this will lead to systematic uncertainties which are difficult to quantify. Some new factorization techniques such as the QCD factorization (QCDF)~\cite{Beneke:1999br,Beneke:2003zv} and the perturbative QCD factorization (pQCD)~\cite{Li:1994iu,Cheng:1999gs} methods etc are employed to solve this problem. In the future works, we can try to use these new techniques to improve the accuracy of analysis for $B_{c}$ two-body weak decays. 

\begin{table*}[htbp]
	\renewcommand\arraystretch{1.5}
	\begin{ruledtabular}
		\caption{Decay widths (in $10^{-7}$eV) and Branching ratios ($10^{-3}$) of $B_c$ nonleptonic decays. The superscript star denotes the results obtained by considering Coulomb-like correction.}
		\label{TBD}
		\begin{tabular}{c >{\centering}p{7em}| c c | c c c c c c c c}
			&\multirow{2}{*}{Decay channels} & \multicolumn{2}{c|}{Decay widths} & \multicolumn{7}{c}{Branching ratios}\\
			&&This work & This work$^*$ &This work & This work$^*$&~\cite{Ebert:2010zu}&~\cite{Hernandez:2006gt}&~\cite{Zhang:2023ypl}&~\cite{Rui:2017pre}&~\cite{Zhu:2017lwi}&~\cite{Kiselev:2001zb} \\
			\hline
			&$B_c^- \to \chi_{c0} \pi^-$ & $3.11^{+0.65}_{-0.58}$ & $25.30^{+3.99}_{-4.07}$ & $0.24^{+0.05}_{-0.04}$ &  $1.96^{+0.21}_{-0.30}$&$0.21$&$0.26$&$0.66$&$1.6$&$6.47$&$9.8$ \\
			&$B_c^- \to \chi_{c0} K^-$ & $0.24^{+0.05}_{-0.04}$ & $1.96^{+0.31}_{-0.31}$ & $0.019^{+0.004}_{-0.004}$ &  $0.15^{+0.03}_{-0.02}$&$0.016$&$0.02$&$0.052$&$0.12$&$0.49$&$-$  \\
			&$B_c^- \to \chi_{c0} \rho^-$ & $8.23^{+1.46}_{-1.67}$ & $66.60^{+8.88}_{-11.31}$ & $0.64^{+0.11}_{-0.13}$  & $5.16^{+0.69}_{-0.87}$&$0.58$&$0.67$&$1.69$&$5.8$&$-$&$33$ \\
			&$B_c^- \to \chi_{c0} K^{*-}$ & $0.43^{+0.08}_{-0.09}$ & $3.49^{+0.46}_{-0.59}$ & $0.033^{+0.006}_{-0.006}$ &  $0.27^{+0.04}_{-0.04}$&$0.04$&$0.037$&$0.096$&$0.33$&$-$&$-$\\
            &$B_c^- \to \chi_{c1} \pi^-$ & $0.14^{+0.06}_{-0.05}$ & $0.91^{+0.31}_{-0.34}$ & $0.011^{+0.005}_{-0.004}$ &  $0.070^{+0.025}_{-0.026}$&$0.2$&$0.0014$&$0.13$&$0.51$&$0.064$&$0.089$ \\
			&$B_c^- \to \chi_{c1} K^-$ & $0.011^{+0.004}_{-0.004}$ & $0.070^{+0.024}_{-0.026}$ & $0.00082^{+0.00038}_{-0.00028}$ &  $0.0054^{+0.0019}_{-0.0020}$&$0.015$&$0.00011$&$0.010$&$0.038$&$0.0049$&$-$  \\
			&$B_c^- \to \chi_{c1} \rho^-$ & $0.88^{+0.23}_{-0.23}$ & $6.74^{+1.05}_{-1.74}$ & $0.068^{+0.018}_{-0.018}$  & $0.52^{+0.08}_{-0.13}$&$0.15$&$0.10$&$0.43$&$2.8$&$-$&$4.6$ \\
			&$B_c^- \to \chi_{c1} K^{*-}$ & $0.050^{+0.017}_{-0.010}$ & $0.41^{+0.07}_{-0.10}$ & $0.0041^{+0.0011}_{-0.0010}$ &  $0.032^{+0.005}_{-0.008}$&$0.01$&$0.0073$&$0.027$&$0.18$&$-$&$-$\\
            &$B_c^- \to h_{c} \pi^-$ & $24.07^{+4.74}_{-3.74}$ & $199.98^{+28.92}_{-25.76}$ & $1.87^{+0.36}_{-0.29}$ &  $15.50^{+2.24}_{-1.99}$&$0.46$&$0.53$&$0.96$&$0.54$&$9.73$&$16$ \\
			&$B_c^- \to h_{c} K^-$ & $1.86^{+0.37}_{-0.29}$ & $15.48^{+2.20}_{-1.59}$ & $0.14^{+0.03}_{-0.02}$ &  $1.20^{+0.17}_{-0.15}$&$0.035$&$0.041$&$0.075$&$0.043$&$0.74$&$-$  \\
			&$B_c^- \to h_{c} \rho^-$ & $61.36^{+12.36}_{-9.43}$ & $523.62^{+65.30}_{-78.40}$ & $4.76^{+0.95}_{-0.73}$  & $40.59^{+5.06}_{-6.08}$&$1.0$&$1.3$&$2.42$&$2.3$&$-$&$53$ \\
			&$B_c^- \to h_{c} K^{*-}$ & $3.44^{+0.68}_{-0.53}$ & $29.20^{+3.64}_{-4.35}$ & $0.27^{+0.05}_{-0.04}$ &  $2.26^{+0.29}_{-0.33}$&$0.07$&$0.071$&$0.13$&$0.13$&$-$&$-$\\
            &$B_c^- \to \chi_{c2} \pi^-$ & $19.43^{+1.94}_{-3.97}$ & $142.31^{+19.87}_{-20.76}$ & $1.51^{+0.15}_{-0.31}$ &  $11.03^{+1.54}_{-1.61}$&$0.38$&$0.22$&$-$&$4.0$&$4.37$&$8.3$ \\
			&$B_c^- \to \chi_{c2} K^-$ & $1.45^{+0.15}_{-0.29}$ & $10.63^{+1.50}_{-1.53}$ & $0.11^{+0.01}_{-0.02}$ &  $0.82^{+0.12}_{-0.11}$&$0.028$&$0.017$&$-$&$0.31$&$0.33$&$-$  \\
			&$B_c^- \to \chi_{c2} \rho^-$ & $52.70^{+5.31}_{-9.90}$ & $387.82^{+54.56}_{-53.93}$ & $4.09^{+0.41}_{-0.77}$  & $30.06^{+4.23}_{-4.18}$&$1.10$&$0.65$&$-$&$16$&$-$&$33$ \\
			&$B_c^- \to \chi_{c2} K^{*-}$ & $2.80^{+0.28}_{-0.51}$ & $20.61^{+2.90}_{-2.82}$ & $0.22^{+0.02}_{-0.04}$ &  $1.60^{+0.22}_{-0.22}$&$0.074$&$0.038$&$-$&$0.96$&$-$&$-$\\
		\end{tabular}
	\end{ruledtabular}
\end{table*}

\section{Conclusions}\label{sec5}
In this work, the vector, axial vector and tensor form factors of $B_{c}$ to $P$-wave charmonia in space-like regions are firstly analyzed in three-point QCD sum rules where the contribution of gluon condensate and Coulomb-like correction are considered. Then, these form factors in zero point and time-like regions are obtained by fitting the results within $z-$ series parameterization approach. With calculated vector and axial vector form factors, we directly analyze the three-body semileptonic decay processes of $B_{c}$ to $P$-wave charmonia. In addition, the two-body nonleptonic decay processes of $B_{c}$ to $P$-wave charmonia and light pseudoscalar or vector mesons are also analyzed by using NFA. We hope these results can help to shed more light on the properties of $B_{c}$ meson and $P$-wave charmonia and provide useful information to research the heavy-flavor physics.

\section*{Acknowledgements}

This project is supported by National Natural Science Foundation under the Grant No. 12175037, No. 12335001, No. 12175068, and Natural Science Foundation of HeBei Province under the Grant No. A2018502124.

\begin{widetext}
\appendix

\section{The graphics of Borel platforms for all form factors.}\label{Sec:AppA}
See Figs. \ref{BW1} and \ref{BW2}.
\begin{figure*}[htbp]
\centering
\includegraphics[width=16.5cm, clip]{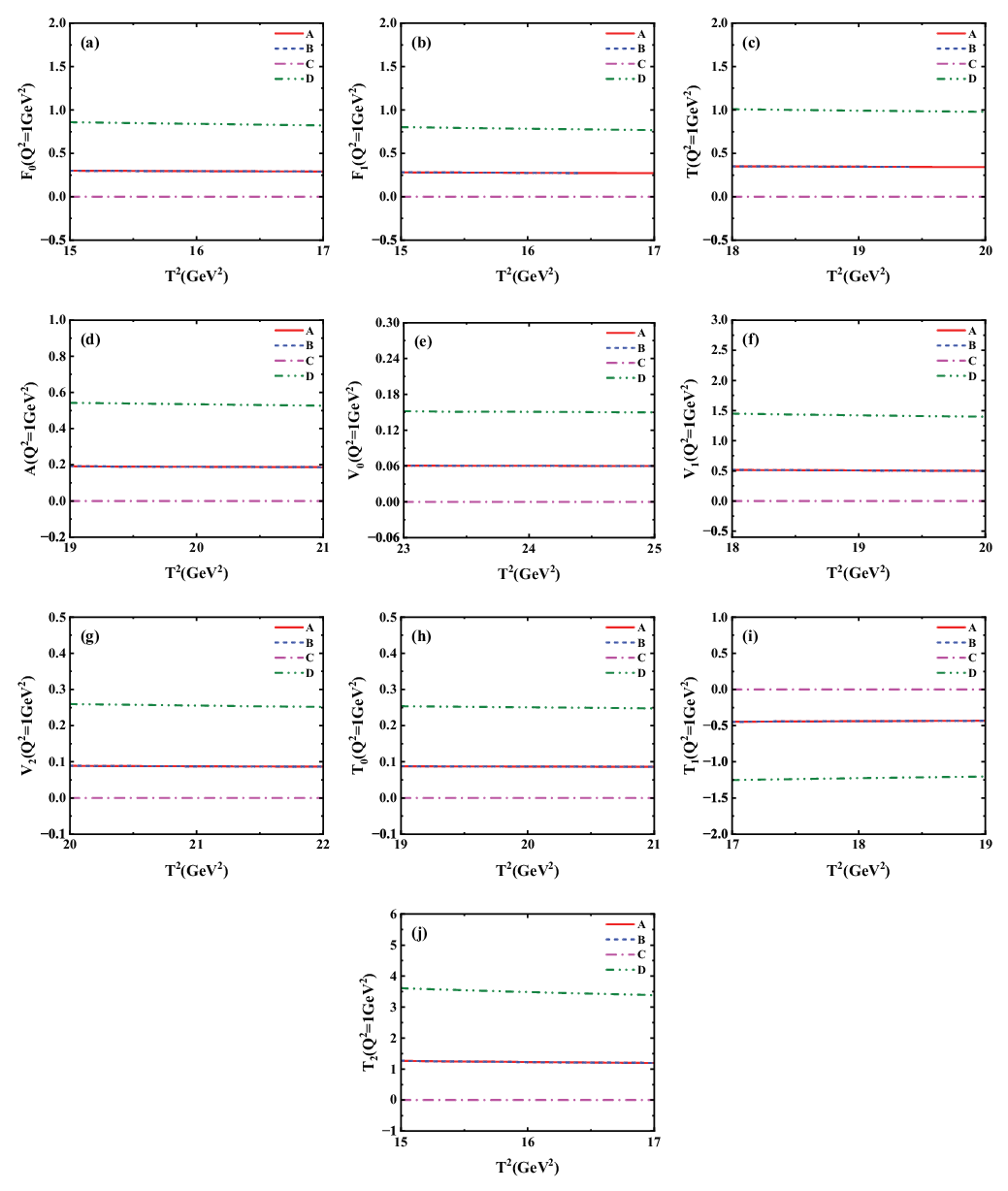}
\caption{The contributions of perturbative and non-perturbative terms with variation of Borel parameter $T^{2}$. (a-c) are for axial vector and tensor form factors of $B_{c}\to \chi_{c0}$, (d-j) are for axial vector, vector and tensor form factors of $B_{c}\to \chi_{c1}$. The legends A-C denote the total, perturbative, gluon condensate contributions, and D represent the total contribution by considering the Coulomb-like correction.}
	\label{BW1}
\end{figure*}
\begin{figure*}[htbp]
\centering
\includegraphics[width=16.5cm, clip]{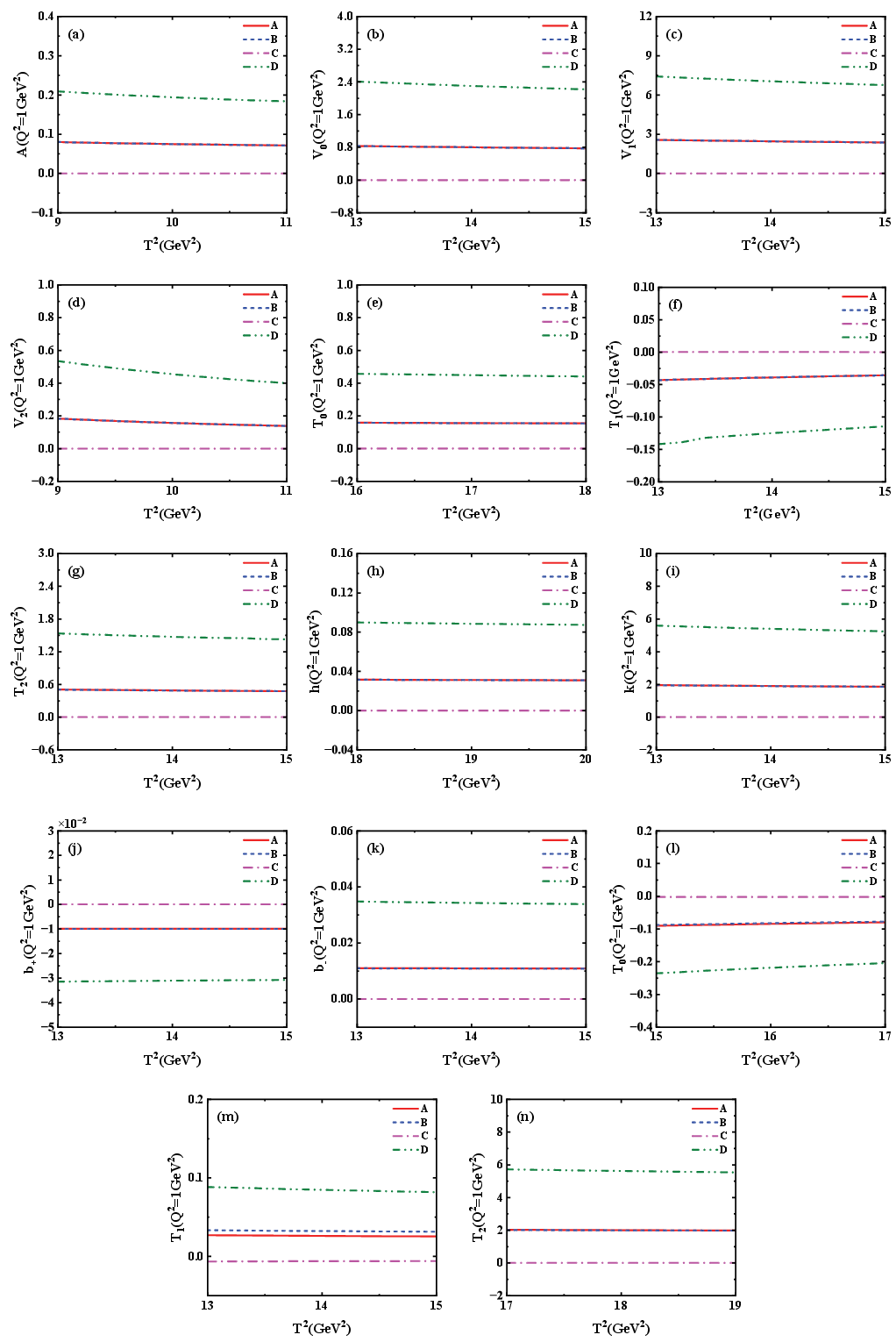}
\caption{The contributions of perturbative and non-perturbative terms with variation of Borel parameter $T^{2}$. (a-g) are for axial vector, vector and tensor form factors of $B_{c}\to h_{c}$ and (h-n) are for axial vector, vector and tensor form factors of $B_{c}\to \chi_{c2}$. The legends are same as Fig. \ref{BW1}.}
	\label{BW2}
\end{figure*}

\section{The full expressions of differential decay widths for $B_{c}$ semileptonic decays.}\label{Sec:AppB}
The differential decay width of $B_{c}\to \chi_{c0}l\bar{\nu}_{l}$ can be expressed as,
\begin{eqnarray}\label{eq:45}
	\notag
	\frac{d\Gamma (B_c \to \chi _{c0}l\bar \nu _l)}{dq^2} &&= \big(\frac{q^2 - m_l^2}{q^2}\big)^2\frac{G_F^2V_{cb}^2\lambda (m_{B_c}^2,m_{\chi _{c0}}^2,q^2)^{1/2}}{384\pi ^3m_{B_c}^3q^2}\left[(m_l^2 + 2q^2)\lambda (m_{B_c}^2,m_{\chi _{c0}}^2,q^2)F_1(q^2)^2 \right.\\
	&&\left. + 3m_l^2(m_{B_c}^2 - m_{\chi _{c0}}^2)^2F_0(q^2)^2 \right]
\end{eqnarray}
where the lambda function can be defined as,
\begin{eqnarray}\label{eq:46}
\lambda(a,b,c)=a^2+b^2+c^2-2(ab+ac+bc)
\end{eqnarray}
The expressions of differential decay width for $B_{c}\to \chi_{c1}[h_{c}]l\bar{\nu}_{l}$ is as follow,
\begin{eqnarray}\label{eq:47}
\frac{d\Gamma (B_c \to \chi _{c1}[h_c]l{\bar \nu _l})}{dq^2} = \frac{d\Gamma ^L(B_c \to \chi _{c1}[h_c]l\bar \nu _l)}{dq^2} + \frac{d\Gamma ^ + (B_c \to \chi _{c1}[h_c]l\bar \nu _l)}{dq^2} + \frac{d\Gamma ^ - (B_c \to \chi _{c1}[h_c]l\bar \nu _l)}{dq^2}
\end{eqnarray}
where,
\begin{eqnarray}\label{eq:48}
	\notag
		\frac{d\Gamma ^L(B_c \to \chi _{c1}[h_c]l\bar \nu _l)}{dq^2} &&= \big(\frac{q^2 - m_l^2}{q^2}\big)^2\frac{G_F^2V_{cb}^2\lambda (m_{B_c}^2,m_{\chi _{c1}[h_c]}^2,q^2)^{1/2}}{384\pi ^3m_{B_c}^3q^2}\left\{ 3m_l^2 \lambda (m_{B_c}^2,m_{\chi _{c1}[h_c]}^2,q^2)V_0(q^2)^2 + \frac{m_l^2 + 2q^2}{4m_{\chi _{c1}[h_c]}^2}\right. \\
		\notag
		&&\times \left. \left. \left[ (m_{B_c}^2 - m_{\chi _{c1}[h_c]}^2 - q^2) \right.(m_{B_c} - m_{\chi _{c1}[h_c]})V_1(q^2)  - \frac{\lambda (m_{B_c}^2,m_{\chi _{c1}[h_c]}^2,q^2)}{m_{B_c} - m_{\chi _{c1}[h_c]}}V_2(q^2) \right]^2 \right\}\\
		\notag
		\frac{d\Gamma ^ \pm (B_c\to \chi _{c1}[h_c]l\bar \nu _l)}{dq^2} &&= \big(\frac{q^2 - m_l^2}{q^2}\big)^2\frac{G_F^2V_{cb}^2\lambda (m_{B_c}^2,m_{\chi _{c1}[h_c]}^2,q^2)^{1/2}}{384\pi ^3m_{B_c}^3}\left\{ (m_l^2 + 2q^2) \lambda (m_{B_c}^2,m_{\chi _{c1}[h_c]}^2,q^2)\right. \\
		&&\times \left. \left[ \frac{A(q^2)}{m_{B_c} - m_{\chi _{c1}[h_c]}}\left.  \mp \frac{(m_{B_c} - m_{\chi _{c1}}[h_c])V_1(q^2)}{\lambda (m_{B_c}^2,m_{\chi _{c1}[h_c]}^2,q^2)^{1/2}} \right] \right.^2 \right\}
\end{eqnarray}
The differential decay width of $B_{c}\to \chi_{c2}l\bar{\nu}_{l}$ can be expressed as,
\begin{eqnarray}\label{eq:49}
	\frac{d\Gamma (B_c \to \chi _{c2}l\bar \nu _l)}{dq^2} =\frac{d\Gamma ^L(B_c \to \chi _{c2}l\bar \nu _l)}{dq^2} + \frac{d\Gamma ^ + (B_c \to \chi _{c2}l\bar \nu _l)}{dq^2}+ \frac{d\Gamma ^ - (B_c \to \chi _{c2}l\bar \nu _l)}{dq^2}
\end{eqnarray}
where,
\begin{eqnarray}\label{eq:50}
	\notag
	\frac{d\Gamma ^L(B_c \to \chi _{c2}l\bar \nu _l)}{dq^2} &&= \big(\frac{q^2 - m_l^2}{q^2}\big)^2\frac{G_F^2V_{cb}^2\lambda (m_{B_c}^2,m_{\chi _{c2}}^2,q^2)^{3/2}}{3072\pi ^3m_{B_c}^3m_{\chi _{c2}}^2q^2}\left\{ 3m_l^2 \lambda (m_{B_c}^2,m_{\chi _{c2}}^2,q^2)V_0(q^2)^2 + \frac{m_l^2 + 2q^2}{4m_{\chi _{c2}}^2}\right.\\
	\notag
	&&\times \left. \left[ (m_{B_c}^2 - m_{\chi _{c2}}^2 - q^2) \right.(m_{B_c} - m_{\chi _{c2}})V_1(q^2)\left. \left. - \frac{\lambda (m_{B_c}^2,m_{\chi _{c2}}^2,q^2)}{m_{B_c} - m_{\chi _{c2}}}V_2(q^2) \right]^2 \right. \right\}\\
	\notag
	\frac{d\Gamma ^ \pm (B_c \to \chi _{c2}l\bar \nu _l)}{dq^2} &&= (\frac{q^2 - m_l^2}{q^2})^2\frac{G_F^2V_{cb}^2\lambda (m_{B_c}^2,m_{\chi _{c2}}^2,q^2)^{3/2}}{2304\pi ^3m_{\chi _{c2}}^2m_{B_c}^3}\left\{ (m_l^2 + 2q^2) \lambda (m_{B_c}^2,m_{\chi _{c2}}^2,q^2)\right.\\
	&&\times \left. \left[ \frac{A(q^2)}{m_{B_c} - m_{\chi _{c2}}}\left. \mp \frac{(m_{B_c} - m_{\chi _{c2}})V_1(q^2)}{\lambda (m_{B_c}^2,m_{\chi _{c2}}^2,q^2)^{1/2}} \right] \right.^2 \right\}
\end{eqnarray}
The definitions of form factors $A$, $V_{0}$, $V_{1}$ and $V_{2}$ for $B_{c}\to \chi_{c2}$ are given by Eq. (\ref{eq:9}). 

\section{The full expressions of decay widths for $B_{c}$ nonleptonic decays.}\label{Sec:AppC}
The full expressions of decay widths for two-body nonleptonic decays $B_{c}\to X P$ and $B_{c}\to X V$ are as follows, 
\begin{eqnarray}\label{eq:51}
	\notag
	\Gamma (B_c \to \chi _{c0}P) &&= \frac{G_F^2f_P^2V_{cb}^2V_{uq}^2\lambda (m_{B_c}^2,m_{\chi _{c0}}^2,m_P^2)^{1/2}}{32\pi m_{B_c}^3}a_1^2F_0(m_P^2)^2 \big(m_{B_c}^2 - m_{\chi _{c0}}^2\big)^2\\
	\Gamma (B_c \to \chi _{c0}V) &&= \frac{G_F^2f_V^2V_{cb}^2V_{uq}^2\lambda(m_{B_c}^2,m_{\chi _{c0}}^2,m_V^2)^{3/2}}{32\pi m_{B_c}^3}a_1^2F_1(m_V^2)^2
\end{eqnarray}

\begin{eqnarray}\label{eq:52}
	\notag
	\Gamma (B_c \to \chi_{c1}[h_c]P) &&= \frac{G_F^2f_P^2V_{cb}^2V_{uq}^2\lambda(m_{B_c}^2,m_{\chi_{c1}[h_c]}^2,m_P^2)^{3/2}}{32\pi m_{B_c}^3}a_1^2 V_0(m_P^2)^2 \\
	\Gamma (B_c \to \chi_{c1}[h_c]V) &&= \Gamma ^L(B_c \to \chi_{c1}[h_c]V)+\Gamma^+(B_c \to \chi_{c1}[h_c]V)+\Gamma^-(B_c \to \chi_{c1}[h_c]V)
\end{eqnarray}
where,
\begin{eqnarray}\label{eq:53}
	\notag
	\Gamma ^L(B_c \to \chi _{c1}[h_c]V) &&= \frac{\lambda (m_{B_c}^2,m_{\chi _{c1}[h_c]}^2,m_V^2)^{1/2}}{32\pi m_{B_c}^3}G_F^2V_{cb}^2V_{uq}^2f_V^2m_V^2a_1^2\left[ (m_{B_c} - m_{\chi _{c1}[h_c]})\frac{m_{B_c}^2 - m_{\chi _{c1}[h_c]}^2 - m_V^2}{2m_{\chi _{c1}[h_c]}m_V}V_1(m_V^2) \right. \\
	\notag
	&&\left. - \frac{\lambda (m_{B_c}^2,m_{\chi _{c1}[h_c]}^2,m_V^2)}{2m_{\chi _{c1}[h_c]}m_V(m_{B_c} - m_{\chi _{c1}}[h_c])}V_2(m_V^2) \right]^2\\
	\notag
	\Gamma ^ \pm (B_c \to \chi _{c1}[h_c]V) &&= \frac{\lambda (m_{B_c}^2,m_{\chi _{c1}[h_c]}^2,m_V^2)^{1/2}}{32\pi m_{B_c}^3}G_F^2V_{cb}^2V_{uq}^2f_V^2m_V^2a_1^2\left[ (m_{B_c} - m_{\chi _{c1}})V_1(m_V^2) \right.\\
	&&\mp \left. \frac{\lambda (m_{B_c}^2,m_{\chi _{c1}[h_c]}^2,m_V^2)^{1/2}}{2(m_{B_c} - m_{\chi _{c1}})}A(m_V^2) \right]^2
\end{eqnarray}
\begin{eqnarray}\label{eq:54}
	\notag
	\Gamma (B_c \to \chi _{c2}P) &&= \frac{G_F^2f_P^2V_{cb}^2V_{uq}^2\lambda (m_{B_c}^2,m_{\chi _{c2}}^2,m_P^2)^{5/2}}{192\pi m_{B_c}^3m_{\chi _{c2}}^2}a_1^2V_0(m_P^2)^2\\
	\Gamma (B_c \to \chi_{c2}V) &&= \Gamma ^L(B_c \to \chi _{c2}V) + \Gamma ^ +(B_c \to \chi _{c2}V) + \Gamma ^ -(B_c \to \chi _{c2}V)
\end{eqnarray}
where,
\begin{eqnarray}\label{eq:55}
	\notag
	\Gamma ^L(B_c \to \chi _{c2}V) &&= \frac{\lambda (m_{B_c}^2,m_{\chi _{c2}}^2,m_V^2)^{3/2}}{192\pi m_{B_c}^3m_{\chi _{c2}}^2}G_F^2V_{cb}^2V_{uq}^2f_V^2m_V^2a_1^2\left[ (m_{B_c} - m_{\chi _{c2}})\frac{m_{B_c}^2 - m_{\chi _{c2}}^2 - m_V^2}{2m_{\chi _{c2}}m_V}V_1(m_V^2) \right.\\
	\notag
	&&\left. - \frac{\lambda (m_{B_c}^2,m_{\chi _{c2}}^2,m_V^2)}{2m_{\chi _{c2}}m_V(m_{B_c} - m_{\chi _{c2}})}V_2(m_V^2) \right]^2\\
	\notag
	\Gamma ^ \pm (B_c \to \chi _{c2}V) &&= \frac{\lambda (m_{B_c}^2,m_{\chi _{c2}}^2,m_P^2)^{3/2}}{256\pi m_{B_c}^3m_{\chi _{c2}}^2}G_F^2V_{cb}^2V_{uq}^2f_V^2m_V^2a_1^2\left[(m_{B_c} - m_{\chi _{c2}})V_1(m_V^2) \right.\\
	&&\mp \left. \frac{\lambda (m_{B_c}^2,m_{\chi _{c2}}^2,m_V^2)^{1/2}}{2(m_{B_c} - m_{\chi _{c2}})}A(m_V^2) \right]^2
\end{eqnarray}
\end{widetext}

\bibliography{ref.bib}

\end{document}